\documentclass[11pt]{amsart}
\usepackage{geometry}                
\usepackage{graphicx}
\usepackage{amsmath}
\usepackage{amssymb}
\usepackage{epstopdf}
\usepackage{amsaddr}

\usepackage{color}
\usepackage{bbm}

\usepackage[tableposition=top]{caption}

\newcommand{\bx}{\mathbf{x}}
\newcommand{\bv}{\mathbf{v}}
\newcommand{\bB}{\mathbf{B}}
\newcommand{\bb}{\mathbf{b}}
\newcommand{\bE}{\mathbf{E}}
\newcommand{\bF}{\mathbf{F}}
\newcommand{\bG}{\mathbf{G}}
\newcommand{\bu}{\mathbf{u}}
\newcommand{\bI}{\mathbf{I}}
\newcommand{\dt}{\Delta t}
\newcommand{\nph}{^{n+1/2}}
\newcommand{\Oc}{\Omega_c}
\newcommand{\deriv}[2]{\frac{d #1}{d #2}}
\newcommand{\vpar}{v_\parallel}
\newcommand{\vtil}{\tilde{\mathbf{v}}}

\title[Conservative AP Integrator]{An energy-conserving and asymptotic-preserving charged-particle orbit implicit time integrator for arbitrary electromagnetic fields}
\author{L.F.\ Ricketson}
\address[L.F.\ Ricketson]{Lawrence Livermore National Laboratory, Livermore, CA 94550, United States}
\email[L.F.\ Ricketson]{ricketson1@llnl.gov}

\author{L.\ Chac\'{o}n}
\address[L.\ Chac\'{o}n]{Los Alamos National Laboratory, Los Alamos, NM 87545, United States}

\begin{document}

\begin{abstract}
We present a new implicit asymptotic preserving time integration scheme for charged-particle orbit computation in arbitrary electromagnetic fields.  The scheme is built on the Crank-Nicolson integrator and continues to recover full-orbit motion in the small time-step limit, but also recovers all the first-order guiding center drifts as well as the correct gyroradius when stepping over the gyration time-scale.  In contrast to previous efforts in this direction, the new scheme also features exact energy conservation.  In the derivation of the scheme, we find that a new numerical time-scale is introduced.  This scale is analyzed and the resulting restrictions on time-step are derived.  Based on this analysis, we develop an adaptive time-stepping strategy the respects these constraints while stepping over the gyration scale when physically justified.  It is shown through numerical tests on single-particle motion that the scheme's energy conservation property results in tremendous improvements in accuracy, and that the scheme is able to transition smoothly between magnetized and unmagnetized regimes as a result of the adaptive time-stepping.  
\end{abstract}
\maketitle

\section{Introduction}

In recent years, considerable progress has been made on implicit particle-in-cell (PIC) methods - e.g. \cite{chacon2013charge, chen2014energy, chen2015multi, chen2011energy, chen2012efficient, chen2014fluid, taitano2013development}. Prominent among the advantages of these methods is their exact total energy conservation for any time-step and cell size. Due to this conservation, implicit PIC schemes have been shown to be more robust against the finite-grid instability than their explicit counterparts. In addition to the obvious benefits of exact energy conservation for long-time accuracy, resistance to this instability facilitates enormous speed-ups when solution structures of interest are much larger than the Debye length.

However, in strongly magnetized plasmas, there is another onerous time-step constraint not overcome by these implicit schemes - namely, the time-step must resolve the gyroperiod. Traditionally, this difficulty has been overcome using gyrokinetic plasma models \cite{chang2004numerical, chen2001gyrokinetic, ethier2005gyrokinetic, FIVAZ199827, Krommes_2007}, in which the gyroperiod time-scale has been analytically eliminated by an asymptotic expansion.

This gyrokinetic approach presents two difficulties. Firstly, the exact energy conservation property enjoyed by implicit schemes is not easily carried over to the gyrokinetic context.  Indeed, some gyrokinetic models do not conserve energy even at the continuum level, and those that do conserve only an approximation of the total full-orbit energy \cite{cary:2009aa}. Secondly, in many problems of interest, the gyrokinetic approximation may be valid only in a subset of the spatial domain. If the transition region - between strongly and weakly magnetized portions of the domain - is narrow and its location is known \textit{a priori}, then a domain decomposition method can be effective. However, in challenging problems of scientific interest, the transition region may be wide and its location may be unknown and/or time varying.

An alternate approach to overcoming the gyroperiod time-step restriction that side- steps these difficulties is to derive an asymptotic preserving (AP) time-stepping scheme. That is, one seeks a time-stepping scheme for particle evolution that (a) recovers the exact dynamics as $\Delta t \rightarrow 0$, (b) recovers the guiding-center particle motion when $\Omega_c \Delta t \gg 1$, where $\Omega_c = qB/m$ is the gyro-frequency, and (c) transitions seamlessly between these two regimes. Several such schemes exist in the literature \cite{brackbill1985simulation, cohen2007large, filbet2016asymptotically, filbet2017asymptotically, genoni2010fast, vu1995accurate}, but none feature exact total energy conservation in the context of a PIC scheme.

The purpose of the present article is to propose a new implicit, AP time-stepping method that exactly conserves total energy. The starting point is the Crank-Nicolson integrator used in prior implicit PIC work - see citations above. Building on asymptotic and empirical
results in \cite{brackbill1985simulation}, we show that, in contrast to the classical Boris scheme (which is known to feature an artificially enlarged gyroradius for $\Omega_c \Delta t \gg 1$ \cite{birdsall2004plasma, parker1991numerical}), Crank-Nicolson recovers the correct gyroradius for arbitrary time-step. We then add modifications that borrow ideas from both \cite{brackbill1985simulation, vu1995accurate} and \cite{genoni2010fast} to ensure that the scheme captures the $\nabla B$ drift while preserving energy conservation. We refer to prior work \cite{parker1991numerical} to find that the scheme captures the inertial drifts for arbitrary time-step with no additional modifications.

Additionally, we derive expressions for the time-step restrictions on our new scheme. It is shown that when the assumptions necessary for drift motion are satisfied, these time- step limits do in fact allow for $\Omega_c \Delta t \gg 1$. We present an adaptive time-stepping scheme that ensures that these new restrictions are obeyed at each time step in spatiotemporally varying fields.

Finally, our scheme is compared against the standard Boris push and existing AP integrators.  This is done for single particle motion in several field configurations of varying complexity.  We find that the enforcement of energy conservation gives the new scheme dramatically improved long-time accuracy compared to previous efforts.  For instance, while adiabatic invariance of magnetic moment is not explicitly enforced, we find that the new scheme conserves it to a good approximation in magnetized regions while previous schemes fail to do so.

The remainder of the article is structured as follows. In Section 2, we review background material, including the requirements for energy conservation in implicit PIC, guiding center motion, properties of Crank-Nicolson applied to charged particle motion, and existing AP integrators. In Section 3, we develop our energy-conserving modification to capture the $\nabla B$ drift; we also derive time-step restrictions for our scheme and present an adaptive time- stepping strategy. We present numerical examples in Section 4 and conclude in Section 5.

\section{Review}

\subsection{Implicit PIC}
At the heart of recent energy-conserving implicit PIC schemes is the Crank-Nicolson integrator for updating particle position and velocity at time $t_n = n \Delta t$, denoted by $(\bx^n_p, \bv_n^p)$, to time $t^{n+1}$:
\begin{equation} \label{eq:CNdef}
\begin{split}
	\bv_p^{n+1} &= \bv_p^n + \dt \frac{q_p}{m_p} \left( \bE_p\nph + \bv_p\nph \times \bB\nph \right), \\
	\bx_p^{n+1} &= \bx_p^n + \dt \bv_p\nph.
\end{split}
\end{equation}
where the quantities at the half-time-step are defined by
\begin{equation}
\begin{split}
	\bv_p\nph &= \left( \bv_p^n + \bv_p^{n+1} \right)/2, \quad \bx_p\nph = \left( \bx_p^n + \bx_p^{n+1} \right), \\
	\bE_p\nph &= \bE \left( \bx_p\nph, t\nph \right), \quad \bB_p\nph = \bB \left( \bx_p\nph, t\nph \right).
\end{split}
\end{equation}
Here, $\bE(\bx,t)$ and $\bB(\bx,t)$ are computed self-consistently using the particle data as inputs.  

Exact energy conservation is shown under various conditions by noting that
\begin{equation} \label{eq:EconsDerivation}
\begin{split}
	\sum_p \frac{m_p}{2} \left\| \bv_p^{n+1} \right\|^2 - \sum_p \frac{m_p}{2} \left\| \bv_p^{n} \right\|^2 &= \frac{1}{2} \sum_p m_p \bv_p\nph \cdot \left( \bv_p^{n+1} - \bv_p^n \right) \\
	&= \frac{\dt}{2} \sum_p q_p \bv_p\nph \cdot \bE\nph.
\end{split}
\end{equation}
From here, the details vary slightly depending on the context (e.g.\ electrostatic \cite{chen2011energy} or Vlasov-Darwin \cite{chen2015multi}), but the common theme is that if (a) $\bE$ is updated using Ampere's law rather than Poisson's equation, and (b) the current density $\mathbf{j}$ is interpolated to the grid using the same shape function that is used to interpolate $\bE$ from the grid to the particles, then one can choose the discretization such that the second line of \eqref{eq:EconsDerivation} is equivalent to the negation of the change in discrete potential energy from step $n$ to $n + 1$. Thus, the total energy is unchanged by a time-step. See \cite{chen2015multi, chen2011energy} for considerably more detail.

For our purposes, the key step in this analysis lies in going from the first to second line in \eqref{eq:EconsDerivation}. There, we have used the fact that the discrete magnetic-field force is necessarily
orthogonal to $\bv_p\nph$. Any additional forces added to the first line of \eqref{eq:CNdef} that are also
orthogonal to $\bv_p\nph$ will thus not break the energy conservation property. This observation
is crucial in the derivation of our energy-conserving AP integrator, as it allows us to work without considering the details of the field solve and its relation to particle data. As such, for the remainder of the article we are safe in dropping the $p$ subscript everywhere and considering a single particle in prescribed electromagnetic fields.

\subsection{Guiding Center Motion and Crank-Nicolson}
Single particle motion in prescribed electromagnetic fields is well-understood in the asymptotic limit in which the gyroperiod $\Omega_c^{-1}$ is small compared to all other time scales. The standard derivation - see e.g.
\cite{hazeltine2018framework, hazeltine2003plasma} - uses the method of averaging to show that, to leading order, the particle velocity is composed of a rapid gyration $\bu$ superimposed on several slower ``drifts":
\begin{equation}
\begin{split}
	\bv &\approx \bu + v_\parallel \bb + \bv_E + \bv_I + \bv_{\nabla B}, \\
	\bu &= u_\perp ( \cos \Omega_c t \mathbf{e}_1 + \sin \Omega_c t \mathbf{e}_2 ), \\
	\bv_E &= \frac{\bE \times \bB}{B^2}, \\
	\bv_I &= \bv_P + \bv_C, \\
	\bv_P &= \frac{\bb}{\Oc} \times \deriv{\bv_E}{t}, \\
	\bv_C &= \frac{\bb}{\Oc} \times \vpar \deriv{\bb}{t}, \\
	\bv_{\nabla B} &= \frac{\bb}{\Oc} \times \frac{\mu}{m} \nabla B.
\end{split}
\end{equation}
Above, $\bb = \bB / B$, $\{ \mathbf{e}_1, \mathbf{e}_2, \bb \}$ form an orthonormal basis for $\mathbb{R}^3$, $\mu = m u_\perp^2/2B$ is the magnetic moment, and the evolution of $u_\perp$ is specified by conservation of $\mu$ - i.e.\ $\dot{\mu} = 0$.  The parallel velocity solves the ODE
\begin{equation} \label{eq:parODE}
	m \deriv{\vpar}{t} = q E_\parallel - \mu \nabla B \cdot \bb - m \deriv{\bv_E}{t} \cdot \bb
\end{equation}
to leading order.  

These drifts are often given names, which we will find convenient to use: $\bv_E$ is called the ``$\bE \times \bB$ drift", $\bv_I$ the ``inertial drift", and $\bv_{\nabla B}$ the ``magnetic drift" or ``$\nabla B$ drift".  The inertial drift is often decomposed into $\bv_P$ - the ``polarization drift" - and $\bv_C$, which we will refer to as the ``curvature drift". Traditionally, the curvature drift is the name for a particular component of $\bv_C$, but it is frequently the most important one, so we will use this name for all of $\bv_C$.

Additionally, the effective force $-\mu \nabla B \cdot \bb$ appearing in \eqref{eq:parODE} is often called the ``mirror force", so named because it gives rise to particle confinement in a magnetic mirror. In the course of the derivation, it has the same origin as the $\nabla B$ drift - namely, an effective force $-\mu \nabla B$ acts on the guiding center. The perpendicular component of this force results in the $\nabla B$ drift, while the parallel component manifests as the mirror force. For this reason, we shall use the term ``$\nabla B$ drift" to refer both to the explicit drift $\bv_{\nabla B}$ and motion induced by the mirror force when no confusion results.

It is our aim to find a time integration scheme that accurately reproduces all of these drifts when $\Oc \dt \gg 1$. It is straightforward to show that the Crank-Nicolson scheme \eqref{eq:CNdef} already captures the $\bE \times \bB$ drift for arbitrary time-step. However, we present the argument in detail here because it establishes notation that we will find useful in our later analysis. It will also lead us to the conclusion that Crank-Nicolson recovers the correct gyroradius for arbitrary $\dt$.

To that end, let us temporarily assume that $\bE$ and $\bB$ are uniform in time and space so that we may drop their $n+1/2$ superscripts.  We define the new discrete velocity 
\begin{equation}
	\vtil^n = \bv^n - \bv_E.
\end{equation}
Elementary manipulation shows that
\begin{equation} \label{eq:vtildeev}
	\vtil\nph = \vtil^n + \frac{\dt}{2} \frac{q}{m} \left( \bE_\parallel + \vtil\nph \times \bB \right).
\end{equation}
We wish to focus on perpendicular motion, so we cross the above with $\bb$ to find
\begin{equation} \label{eq:crosswb}
	\vtil\nph \times \bb = \vtil^n \times \bb - \frac{\dt}{2} \frac{q}{m} \vtil\nph_\perp B.
\end{equation}
Substituting this into the perpendicular component of \eqref{eq:vtildeev} gives
\begin{equation} \label{eq:vnphrot}
	\vtil\nph_\perp = \frac{ \vtil_\perp^n + \frac{1}{2} \Oc \dt \vtil^n \times \bb }{ 1 + \Oc^2 \dt^2 / 4 }.
\end{equation}
From here, simple manipulation gives
\begin{equation} \label{eq:RthetaDef}
	\vtil^{n+1}_\perp = \frac{ \left( 1 - \Oc^2 \dt^2 / 4 \right) \vtil^n_\perp + \Oc \dt \left( \vtil^n \times \bb \right) }{1 + \Oc^2 \dt^2 / 4}.
\end{equation}
Dotting both sides with themselves and doing some algebra, we can see that $\left\| \vtil_\perp^{n+1} \right\| = \left\| \vtil^n_\perp \right\|$.  This reveals that the transformation $\vtil^n \rightarrow \vtil^{n+1}$ is a rotation.  Dotting with $\vtil^n$ shows that the rotation angle $\theta$ satisfies
\begin{equation} \label{eq:thetaDef}
	\cos \theta = \frac{ 1 - \Oc^2 \dt^2 / 4}{ 1 + \Oc^2 \dt^2 / 4}.
\end{equation}

In constant, uniform fields, we can thus write the perpendicular component of the velocity update as
\begin{equation} \label{eq:constfieldrot}
	\bv_\perp^{n+1} = \bv_E + R_\theta \left[ \bv_\perp^n - \bv_E \right],
\end{equation}
where $R_\theta$ denotes the rotation matrix about $\bb$ by angle $\theta$ given above - i.e.\ it can be defined by the relation $\vtil_\perp^{n+1} = R_\theta \vtil_\perp^n$ along with \eqref{eq:RthetaDef}.  In uniform fields, this would be the \textit{exact} solution if it were the case that $\theta = \Oc \dt$.  However, $\theta$ only approximates $\Oc \dt$ for $\Oc \dt \ll 1$, as can be easily seen from Taylor expansion of the right side of \eqref{eq:thetaDef}.  It bears noting that for $\Oc \dt \gg 1$, the rotation angle $\theta$ tends to $\pi$.  This observation lends intuition to later analysis.  

Thus, we see that at least in constant, homogeneous fields, Crank-Nicolson exactly recovers the correct $\bE \times \bB$ drift, but makes errors in the gyro-phase of the particle that grow with time-step. However, in the strongly magnetized case, errors in gyro-phase are of little concern, since we assume that the gyroradius is small compared to length scales of interest.

Finally, we note that nothing in this derivation actually requires $\bE$ and $\bB$ to be spatially or temporally uniform - it was simply convenient to drop the superscripts.  We may thus reintroduce the $n+1/2$ superscripts on all fields now.  The only additional subtlety is that the $\perp$ component is always with respect to $\bb^{n+1/2}$, regardless of the time-level at which we are evaluating the velocity - this is a consequence of \eqref{eq:crosswb}, in which the $\bb$ in the cross product is evaluated at $t\nph$.  The general form of \eqref{eq:constfieldrot} is thus
\begin{equation}
	\left( \bI - \bb\nph \bb\nph \right) \cdot \bv^{n+1} = \bv_E\nph + R_\theta\nph \left[ \left( \bI - \bb\nph \bb\nph \right) \cdot \bv^n - \bv_E\nph \right],
\end{equation}
where $\bI$ is the identity tensor.  Thus, even in varying fields, the $\bE \times \bB$ drift is approximately recovered so long as $\bb$ and $\bv_E$ change little within a time-step.  Since each is assumed to vary on a scale much longer than the gyro-period, this is a much less restrictive constraint than $\Oc \dt \ll 1$.  

\subsection{Numerical Gyroradius}
It is well known that the classical, explicit Boris time integrator, which may be defined by
\begin{equation}
\begin{split}
	\bx\nph &= \bx^{n-1/2} + \dt \bv^n, \\
	\bv^{n+1} &= \bv^n + \dt \frac{q}{m} \left( \bE\nph + \bv\nph \times \bB\nph \right)
\end{split}
\end{equation}
features a numerical gyroradius much larger than the true gyroradius for $\Oc \dt \gg 1$ \cite{birdsall2004plasma, parker1991numerical}.  In particular, the velocity update is identical to the Crank-Nicolson scheme defined above, so in a homogeneous magnetic field with $\bE = 0$, the velocity update still represents rotation by the angle $\theta$ defined in \eqref{eq:thetaDef}.  Geometric arguments show that when the position is updated using a velocity of constant magnitude $v_\perp$ rotating at each time step by an angle $\phi$, the radius of the circular motion is given by \cite{birdsall2004plasma, parker1991numerical} (see particularly Figure 4-3b in \cite{birdsall2004plasma})
\begin{equation} \label{eq:discreteradius}
	\rho_\textrm{eff} = \frac{v_\perp \dt}{2 \left\lvert \sin \left( \phi / 2 \right) \right\rvert}.
\end{equation}
In the Boris push, the perpendicular velocity used to update position is simply $\bv_\perp^n$.  Making use of the trigonometric identity $\left\lvert \sin \left( \phi / 2 \right) \right\rvert = \sqrt{(1 - \cos \phi)/2}$ and \eqref{eq:thetaDef}, one finds that
\begin{equation} \label{eq:rhoBoris}
	\rho^\textrm{Boris}_\textrm{eff} = \rho_\textrm{true} \sqrt{1 + \frac{\Oc^2 \dt^2}{4}},
\end{equation}
where $\rho_\textrm{true} = v_\perp^n / \Oc$ is the physically correct gyroradius.  This clearly has negative consequences for accuracy in spatially varying fields when $\Oc \dt \gg 1$, as discussed in \cite{vu1995accurate}.  

In contrast, when using the Crank-Nicolson scheme defined in \eqref{eq:CNdef}, the velocity used in updating position is not $\bv^n$ but $\bv^{n+1/2}$.  By returning to \eqref{eq:vnphrot} and dotting $\bv_\perp\nph$ with itself (we may drop the tilde since we are assuming $\bE = 0$), we find
\begin{equation} \label{eq:vnphsmall}
	v_\perp\nph = \frac{v_\perp^n}{\sqrt{1 + \Oc^2 \dt^2/4}}.
\end{equation}
This statement may initially be counter-intuitive, but may be explained in the following way: We have already established that for $\Oc \dt \gg 1$, the angle $\theta$ through with $\bv_\perp$ rotates in a time step is nearly $\pi$, so that $\bv_\perp^{n+1} \approx -\bv_\perp^n$.  Thus, $\bv_\perp\nph$, being an average of two vectors which are approximately the negations of each other, should be much smaller than either.  Indeed, $v_\perp\nph$ should tend to zero as $\Oc \dt$ diverges, which is precisely what \eqref{eq:vnphsmall} states.

Substituting this expression for $\bv_\perp$ into \eqref{eq:discreteradius}, we find that when the Crank-Nicolson scheme is used, we have
\begin{equation} \label{eq:rhoCN}
	\rho^\textrm{CN}_\textrm{eff} = \rho_\textrm{true}.
\end{equation}
Thus, remarkably, the Crank-Nicolson scheme recovers the \textit{exact} correct gyroradius for \textit{any} value of $\Oc \dt$.  This was shown in the limit $\Oc \dt \gg 1$ in \cite{vu1995accurate}, and observed empirically in the general case in the numerical experiments of \cite{brackbill1985simulation, vu1995accurate}.  Thus, the implicit Crank-Nicolson scheme begins at an advantage over the Boris push when investigating particle dynamics with $\Oc \dt \gg 1$.  

\subsection{Preservation of Inertial Drifts and Non-Preservation of $\nabla B$ Drift}
In \cite{parker1991numerical}, Parker and Birdsall show that the Boris push recovers \textit{all} first order drift motions for arbitrary $\Oc \dt$, albeit with an artificially enlarged gyroradius.  The discretization is shown to only introduce an additional force of $O(q \rho_\textrm{eff}^2 \| D^2 \bB \|/m)$, where $\| D^2 \bB \|$ is a norm of the Hessian of $\bB$.  That is, if the effective gyroradius is small compared to the scale over which $\bB$ varies, this force is small.  

In Parker and Birdsall's numerical guiding center equation derivation, the numerical $\nabla B$ drift arises from an effective force $-\mu_\textrm{eff} \nabla B$, where
\begin{equation}
	\mu_\textrm{eff} = \frac{q \rho_\textrm{eff}^2}{2 \dt} \sin \theta.
\end{equation}
One may use the expressions above for $\theta$ and $\rho_\textrm{eff}^\textrm{Boris}$ to show that, for the Boris push, $\mu_\textrm{eff}^\textrm{Boris} = \mu_\textrm{true} = m u_\perp^2/2B$.  

However, we have just shown that, for Crank-Nicolson, $\rho_\textrm{eff}$ is much smaller than for Boris when $\Oc \dt \gg 1$.  Thus, Crank-Nicolson dramatically underestimates the $\nabla B$ drift for $\Oc \dt \gg 1$.  To be precise, since $\rho_\textrm{eff}^\textrm{CN} = \rho_\textrm{eff}^\textrm{Boris} / \sqrt{ 1 + \Oc^2 \dt^2 / 4 }$ - see \eqref{eq:rhoBoris} and \eqref{eq:rhoCN} - and the rotation angles $\theta$ are identical, we have 
\begin{equation} \label{eq:mueffCN}
	\mu_\textrm{eff}^\textrm{CN} = \frac{\mu_\textrm{true}}{1 + \Oc^2 \dt^2 / 4}.
\end{equation}

This, though, is the only modification of the Parker/Birdsall analysis.  Thus, Crank-Nicolson still captures the inertial drifts - this is verified empirically in our numerical experiments in Section 4.  For this reason, we follow the lead of previous efforts toward AP schemes by focusing on modifying Crank-Nicolson to capture the $\nabla B$ drift.  

\subsection{Prior AP Schemes}
Previous works have succeeded in capturing the $\nabla B$ drift, although without the energy conservation property we seek.  These schemes can be sorted into three categories.  In the first category are schemes \cite{brackbill1985simulation, genoni2010fast, vu1995accurate} that modify the velocity update equation by adding an ``effective" force that approximates $-\mu \nabla B$ for $\Oc \dt \gg 1$ and is negligible for $\Oc \dt \ll 1$.  In the second category are schemes \cite{cohen2007large, genoni2010fast} that modify the position update equation by adding an ``effective" velocity that approximates $\bv_{\nabla B}$ for $\Oc \dt \gg 1$ and is negligible for $\Oc \dt \ll 1$.  In the third category are the semi-implicit Runge-Kutta schemes of \cite{filbet2017asymptotically} - these schemes are restricted to magnetic fields with fixed direction but variable magnitude.

Note that the scheme of Genoni \textit{et al.} \cite{genoni2010fast}, called the Magnetized Implicit (MI) scheme, falls into both of the first two categories in the sense that it uses an effective force to capture the mirror force ($-\mu \nabla \bB \cdot \bb$) and an effective velocity to capture the perpendicular drift $\bv_{\nabla B}$.  

In summarizing previous efforts, we will focus on the first category because it is more compatible with energy conservation than the second category - we make the reasons for this clear in Section 3 - and applicable to more general field geometries than the third category.

\subsubsection{Brackbill-Forslund-Vu}
The scheme developed by Brackbill, Forslund, and Vu in \cite{brackbill1985simulation, vu1995accurate} (denoted BFV henceforth) introduces an effective force $\bF_{BFV}$ defined by
\begin{equation} \label{eq:FBFVdef}
	\bF_{BFV}\nph = -\tilde{\mu} \nabla B, \qquad \tilde{\mu} = m \frac{ \left\| \bv_\perp^{n+1} - \bv_\perp^n \right\|^2}{8B},
\end{equation}
where all quantities without superscripts are evaluated at $t\nph$.  Using the notation in Section 2.2, straightforward manipulation reveals that
\begin{equation} \label{eq:mutildederivation}
\begin{split}
	\left\| \bv_\perp^{n+1} - \bv_\perp^n \right\|^2 &= \left\| \left( \bv_\perp^{n+1} - \bv_E\nph \right) - \left( \bv_\perp^n - \bv_E^{n+1/2} \right) \right\|^2, \\
	&= \left\| \left[ R_\theta\nph - \bI \right] \left( \bv_\perp^n - \bv_E\nph \right) \right\|^2, \\
	&= \left\| \bv_\perp^n - \bv_E\nph \right\|^2 \frac{\Oc^2 \dt^2}{1 + \Oc^2 \dt^2 / 4}.
\end{split}
\end{equation}
From here, one can read off the following limits:
\begin{equation}
\begin{split}
	\Oc \dt \ll 1 &\implies \tilde{\mu} = O(\dt^2), \\
	\Oc \dt \gg 1 &\implies \tilde{\mu} \approx \frac{m \left\| \bv_\perp^n - \bv_E\nph \right\|^2}{2B}.
\end{split}
\end{equation}
If $\bv_E$ changes little in a time step, then $\left\| \bv_\perp^n - \bv_E\nph \right\| \approx u_\perp^n$, since $\bu$ and $\bv_E$ dominate other drifts in the guiding center limit.  Thus, $\tilde{\mu} \approx \mu$ for $\Oc \dt \gg 1$, and $\bF_{BFV}$ has the features necessary to capture the $\nabla B$ drift for large $\Oc \dt$.  

We can actually go a step further than this.  Recall that applying the analysis in \cite{parker1991numerical} to Crank-Nicolson, we were able to show that unmodified Crank-Nicolson features an effective $\nabla B$ force of $-\mu_\textrm{eff}^\textrm{CN} \nabla B$, with $\mu_\textrm{eff}^\textrm{CN}$ given in \eqref{eq:mueffCN}.  The BFV scheme proposes to modify Crank-Nicolson so that the total force due to $\nabla B$ is 
\begin{equation}
\begin{split}
	- \left( \mu_\textrm{eff}^\textrm{CN} + \tilde{\mu} \right) \nabla B &= - \mu_\textrm{true} \nabla B \left\{ \frac{1}{1 + \Oc^2 \dt^2 / 4} + \frac{\Oc^2 \dt^2 / 4}{1 + \Oc^2 \dt^2 / 4} \right\} \\
	&= -\mu_\textrm{true} \nabla B.
\end{split}
\end{equation}
In the first equality, we have just used \eqref{eq:mutildederivation} and retained our assumption that the gyration velocity is approximated $\bv_\perp - \bv_E$.  Thus, remarkably, the BFV scheme recovers - to leading order - the correct $\nabla B$ force not merely in the limits $\Oc \dt \ll 1$ and $\Oc \dt \gg 1$, but also for \textit{every} value of $\Oc \dt$!

This scheme has two noteworthy drawbacks.  First, and most important for us, is that it does not satisfy the constraint $\bv\nph \cdot \bF_{BFV} = 0$, so it does not conserve energy.  Second, it requires the direct computation of $\nabla B$ at every time-step - in fact, since an iterative solver is typically used to compute the implicit particle update, it must be computed many times per step.

\subsubsection{Magnetized Implicit Scheme}
The MI scheme of Genoni \textit{et al.}\ \cite{genoni2010fast} introduces the effective force 
\begin{equation}
	\bF_{MI}\nph = \frac{q}{4} \left( \bv_\perp^{n+1} - \bv_\perp^n \right) \times \left( \bB^{n+1} - \bB^n \right).
\end{equation}
Strictly speaking, MI uses only the parallel component, $\left( \bF_{MI}\nph \cdot \bb\nph \right) \bb\nph$ - the perpendicular drift is captured by an effective velocity instead.  However, we show that the full $\bF_{MI}$ approximates all components of $-\mu\nabla B$ - note that this is not shown in \cite{genoni2010fast}.  

Note that, using results from Section 2.2, we may rewrite the first term in the cross product defining $\bF_{MI}$ as
\begin{equation}
	\bv_\perp^{n+1} - \bv_\perp^n = \Oc \dt \left( \bv_\perp\nph - \bv_E\nph \right) \times \bb\nph.
\end{equation}
By Taylor expansion, the second term in the cross product may be approximated by
\begin{equation}
\begin{split}
	\bB^{n+1} - \bB^n &= \left( \bx^{n+1} - \bx^n \right) \cdot \nabla \bB\nph + O\left( \left\| \bx^{n+1} - \bx^n \right\|^3 \right) \\
	&= \dt \left( \bv\nph \cdot \nabla \right) \bB\nph + O\left( \dt^3 \right).
\end{split}
\end{equation}
Thus, we can write 
\begin{equation}
	\frac{1}{4} \left( \bv_\perp^{n+1} - \bv_\perp^n \right) \times \left( \bB^{n+1} - \bB^n \right) \approx \frac{1}{4} \Oc \dt^2 \left[ \left( \bv_\perp - \bv_E \right) \times \bb \right] \times \left( \bv \cdot \nabla \right) \bB,
\end{equation}
where all quantities without superscripts are evaluated at $t\nph$.  

Let us now consider the gyroaverage, denoted by $\langle \cdot \rangle$, of this expression.  To leading order, we have
\begin{equation}
	\left\langle \left[ \left( \bv_\perp - \bv_E \right) \times \bb \right] \times \left( \bv \cdot \nabla \right) \bB \right\rangle \approx \left\langle \left( \vtil_\perp \times \bb_{gc} \right) \times \left( \vtil_\perp \cdot \nabla \right) \bB_{gc} \right\rangle,
\end{equation}
where $\bB_{gc}$ denotes the field at the guiding center position, we assumed that $\bB$, the drift velocities, and parallel motion change little in a gyroperiod, and the notation $\vtil = \bv - \bv_E$ has been reintroduced.  We may then follow the standard $\nabla B$ drift derivation in \cite{hazeltine2018framework, hazeltine2003plasma} - namely, expand the triple product and note that $\left\langle \vtil_\perp \vtil_\perp \right\rangle \approx \left\langle \bu \bu \right\rangle = u_\perp^2 ( \bI - \bb \bb)/2$ - to arrive at
\begin{equation}
	\frac{q}{4} \left\langle \left( \bv_\perp^{n+1} - \bv_\perp^n \right) \times \left( \bB^{n+1} - \bB^n \right) \right\rangle \approx - \frac{\Oc^2 \dt^2}{4} \frac{q \left\| \vtil\nph_\perp \right\|^2}{2 \Oc} \nabla B\nph.
\end{equation}
By once again using \eqref{eq:vnphsmall}, we find
\begin{equation}
	\frac{q}{4} \left\langle \left( \bv_\perp^{n+1} - \bv_\perp^n \right) \times \left( \bB^{n+1} - \bB^n \right) \right\rangle \approx -\frac{\Oc^2 \dt^2 / 4}{1 + \Oc^2 \dt^2 / 4} \frac{m \left\| \vtil^n_\perp \right\|^2}{2B} \nabla B\nph.
\end{equation}
This clearly tends to $-\mu \nabla B$ for large $\Oc \dt$, and is $O(\dt^2)$ for $\Oc \dt \ll 1$, as desired.  Additionally, note that similarly to the BFV scheme, this effective force has the property that
\begin{equation}
	-\mu_\textrm{eff}^\textrm{CN} \nabla B + \left\langle \bF_{MI} \right\rangle = -\mu_\textrm{true} \nabla B
\end{equation}
to leading order.  Thus, at least in a gyroaveraged sense, the MI scheme recovers the correct leading order $\nabla B$ force for \textit{every} value of $\Oc \dt$.  

A key difference, of course, between $\bF_{MI}$ and $\bF_{BFV}$ is that the former \textit{only} approximates the correct $\nabla B$ force in a gyroaveraged sense.  The MI scheme thus must rely on the time-stepping process to, in some sense, perform the gyro-averaging implicitly by sampling the particle at different gyro-phases at different time-steps.  This works well in practice for moderate $\Oc \dt$ - see \cite{genoni2010fast} and our numerical results below - but introduces anomalous drifts for very large $\Oc \dt$.  

This limitation will be analyzed in detail in the context of our scheme in the following section, but may be intuitively understood as follows: For very large $\Oc \dt$, the gyration velocity rotates by approximately $\pi$ at every step. It thus takes many time steps to accumulate a representative sample of gyrophases. Before that number of steps is reached, a biased estimate of the gyroaverage is implicitly computed, which gives rise to anomalous displacements. At smaller time steps, an accurate gyroaverage is computed in just a few time steps, and the size of the anomalous displacements is reduced so as to be smaller than the gyroradius.

Of course, this effective force also does not conserve energy. In the full MI scheme, a modification to the perpendicular motion is made to compensate for this, so that it remains true that the magnetic field does no work. Even with that correction, though, the inclusion of an effective velocity to capture $\bv_{\nabla B}$ prevents this scheme from conserving even individual particle energy, much less total energy in the context of an implicit PIC scheme.

This scheme has two benefits.   First, it does not require an explicit computation of $\nabla B$.  Second, as it is presented in \cite{genoni2010fast}, it is actually an \textit{explicit} scheme, using a two-stage predictor-corrector procedure with the quantities evaluated at the $t^{n+1}$ using the values from the predictor stage.  This has obvious benefits for computation speed and implementation simplicity, but further reduces accuracy and conservation as seen in our numerical tests.  

\section{The $\nabla B$ Drift and Energy Conservation}
In seeking to preserve total energy conservation in the context of implicit PIC simulation, we immediately rule out the use of an effective velocity, favoring instead an effective force. This is because it is clear that an effective velocity is likely to break energy conservation. We see this from the following logic: Denote a particle's change in kinetic energy in a single time step by $\Delta K_p$, and its change in potential energy by $\Delta P_p$. The energy conservation property we seek is written  $\sum_p(\Delta K_p + \Delta P_p) = 0$. The velocity update is ignorant of modifications to the position update to leading order, so the addition of an effective velocity will not alter $\Delta K_p$, but will alter $\Delta P_p$ (the particle's position, and thus electrostatic potential energy, is modified).

One has very little hope of maintaining conservation when $\Delta P$ is modified while $\Delta K$ is fixed. Indeed, each particle's new effective velocity would have to conspire to modify its potential energy in such a way that the sum over all modifications is zero - and this must be true for arbitrary field configurations! This circumstance is made even more unlikely by the fact that the effective velocity is meant to capture the $\nabla B$ drift, which is agnostic with respect to the electric field. Altering $\bE$ thus should not alter the effective velocity, meaning it must surely alter $\Delta P$.

In contrast, the addition of an effective force modifies both the position and velocity updates in a self-consistent manner - the position is modified indirectly by the modification of the velocity. Aside from the constraint mentioned above that the new force be orthogonal to $\bv\nph$, the only other place in the implicit PIC energy conservation proofs of \cite{chen2015multi, chen2011energy} that the particle velocity arises is in the computation of the current density. All that is relied upon there is that the position and velocity are related by the Crank-Nicolson position update, which is unmodified by an effective force. Thus, we see that the introduction of an effective force holds the promise of energy conservation (it conserves energy if and only if it is orthogonal to $\bv\nph$), while the introduction of an effective velocity is extremely unlikely to do so.

\subsection{Conservative Effective Force}
Our guiding principle for finding an effective force that both approximates $-\mu \nabla B$ and is orthogonal to $\bv\nph$ is as follows: the effective force is postulated to be the projection of some vector $\bG$ onto the orthogonal complement of $\bv\nph$.  That is,
\begin{equation} \label{eq:simpProjec}
	\bF_{cons} = \left( \bI - \frac{\bv\nph \bv\nph}{\left\| \bv\nph \right\|^2} \right) \cdot \bG,
\end{equation}
for some suitably chosen $\bG$.  An intuitively reasonable choice for $\bG$ would seem to be a scalar multiple of either $\bF_{BFV}$ or $\bF_{MI}$.  This will turn out to be true in certain cases, which we show below.  

This simplistic view, unfortunately, has the problem that it artificially mixes parallel and perpendicular motion. Consider particle motion in the fields $\bE = 0$, $\bB = B(x, y)\widehat{\mathbf{z}}$. In reality, a particle's velocity in the $z$-direction is constant in these fields, since $\nabla B \cdot \bb = 0$. However, the effective force in \eqref{eq:simpProjec} will in general have a non-zero component in the $z$-direction. In particular, the $z$-component is non-zero if the particle initially has non-zero $v_z$ and $\bv\nph$ is not orthogonal to $\bG$. Thus, introduction of this force will modify the parallel velocity even when parallel velocity is physically constant.

This issue is fixed by breaking our projection into parallel and perpendicular components. Instead of \eqref{eq:simpProjec}, we use the modified projection
\begin{equation} \label{eq:goodProjec}
	\bF_{cons} = \left[ \bb - \frac{v_\parallel}{v_\perp} \frac{\bv_\perp}{v_\perp} \right] G_\parallel + \left[ \bI - \frac{\bv_\perp \bv_\perp}{v_\perp^2} \right] \cdot \bG_\perp,
\end{equation}
where all instances of $\bb$ and $\bv$ are evaluated at $t\nph$, and the $\perp$ and $\parallel$ symbols refer to components orthogonal and parallel, respectively, to $\bb\nph$.  Note that each of the two terms appearing in \eqref{eq:goodProjec} is independently orthogonal to $\bv\nph$.  The first term is responsible for the mirror force - i.e.\ $-\mu \nabla B \cdot \bb$ - with a modification to the perpendicular motion to conserve energy, and the second is intended to give rise to $\bv_{\nabla B}$.  In the example above in which parallel velocity should be constant, this effective force behaves correctly so long as $G_\parallel = 0$, since this implies $F_{cons,\parallel} = 0$.  

We now return to our goal, which is that $\bF_{cons}$ should, in some sense, approximate $-\mu \nabla B$ when $\Oc \dt \gg 1$. One consequence of this goal is that we require $\bF_{cons}$ to be independent of gyrophase. However, we are also requiring $\bF_{cons,\perp}$ to be orthogonal to $\bv_\perp$, which manifestly does depend on gyrophase. It is thus unrealistic to expect $\bF_{cons}$ to approximate the $\nabla B$ force at every time-step.

We can, however, enforce that $\bF_{cons}$ approximate the correct $\nabla B$ force in a gyro-averaged sense, as in the MI scheme.  That is, we will specify the vector $\bG$ by insisting that 
\begin{equation}
	\left\langle \bF_{cons} \right\rangle = \bF_{BFV}
\end{equation}
and requiring additionally that $\bG$ be independent of gyrophase.  Note that we have chosen the gyroaverage to equal $\bF_{BFV}$ so that we retain the desirable property enjoyed by both the BFV and MI schemes that the $\nabla B$ force is recovered correctly for \textit{every} value of $\Oc \dt$.  We have chosen $\bF_{BFV}$ instead of $\bF_{MI}$ because the latter is dependent on gyrophase, and the gyroaverages of the two are identical to leading order as shown above.  

It remains only to compute $\bG$ given the constraints above.  We will assume that, as in the particle drift derivation, $\bv$ is dominated jointly by gyration $\bu$, the $\bE \times \bB$ drift $\bv_E$, and parallel velocity $v_\parallel$.  We write $\bv_\perp = \bu + \bv_E$ to leading order, and assume that $v_\parallel$, $\bv_E$, and $\bb$ all change on time-scales much longer that $\Oc^{-1}$.  We can then write the gyroaverage condition above as
\begin{equation} \label{eq:Gconstraint}
\begin{split}
	\bF_{BFV} = &\bb G_\parallel - v_\parallel G_\parallel \left\{ \bv_E \left\langle \frac{1}{\left\| \bu + \bv_E \right\|^2} \right\rangle + \left\langle \frac{\bu}{\left\| \bu + \bv_E \right\|^2} \right\rangle \right\} \\
	&+ \bG_\perp - \left\langle \frac{\bu \bu}{\left\| \bu + \bv_E \right\|^2} \right\rangle \cdot \bG_\perp \\
	&- \left\langle \frac{\bu}{\left\| \bu + \bv_E \right\|^2} \right\rangle \left( \bv_E \cdot \bG_\perp \right) - \bv_E \left( \left\langle \frac{\bu}{\left\| \bu + \bv_E \right\|^2} \right\rangle \cdot \bG_\perp \right) \\
	&- \left\langle \frac{1}{\left\| \bu + \bv_E \right\|^2} \right\rangle \bv_E \left( \bv_E \cdot \bG_\perp \right).
\end{split}
\end{equation}
Only one of the terms on the right is parallel to $\bb$, so we immediately find that $G_\parallel = F_{BFV,\parallel}$.  Note that this does recover constant parallel velocity in the simple case described above.  

By evaluating the remaining three distinct gyroaveraged expressions - see Appendix A for more detail - and substituting in the known value of $G_\parallel$, one can simplify the perpendicular components of the expression above to find
\begin{equation} \label{eq:genConstraint}
	\bF_{BFV,\perp} = -\frac{1 - \eta^2}{\left\lvert 1 - v_E^2/u^2 \right\rvert} \widehat{\bv}_E \left( \frac{v_\parallel v_E}{u^2} F_{BFV,\parallel} + \frac{v_E^2}{u^2} \left( 1 - \eta^2 \right) \widehat{\bv}_E \cdot \bG_\perp \right) + \left( 1 - \frac{\eta^2}{2} \right) \bG_\perp,
\end{equation}
where $\eta = \min \{ 1, u/v_E \}$ and we have introduced $\widehat{\bv}_E = \bv_E/v_E$ for brevity.  Note that it is not surprising that the cases $u > v_E$ and $u < v_E$ are qualitatively different.  Indeed, for $u > v_E$, $\bv_\perp / v_\perp$ traverses the full unit circle in the course of a gyration, while in the oposite case it has a positive component in the $\bv_E$ direction for every gyrophase.  

It is thus sensible to break further analysis into cases: $u>v_E$ (i.e.\ $\eta = 1$) and $u<v_E$ (i.e.\ $\eta = u/v_E < 1$).  In the former case, the entire first term vanishes and we find
\begin{equation} \label{eq:Gperpudom}
	\bG_\perp = 2 \bF_{BFV,\perp} \quad \textrm{if } u > v_E.
\end{equation}
Again, this is unsurprising because $\bv_\perp / v_\perp$ traverses the full unit circle in this case, so the average of the projection operator onto its orthogonal complement is $\bI/2$.  

The case $u < v_E$ is more algebraically cumbersome.  After some simplification - in particular, noting that $v_E / u = \eta^{-1}$ in this case - we can simplify \eqref{eq:genConstraint} to
\begin{equation}
	\bF_{BFV,\perp} = \left( 1 - \frac{\eta^2}{2} \right) \bG_\perp - \left[ \left( 1 - \eta^2 \right) \left( \widehat{\bv}_E \cdot \bG_\perp \right) + F_{BFV,\parallel} \frac{v_\parallel}{v_E} \right] \widehat{\bv}_E.
\end{equation}
This vector equation can be solved for $\bG_\perp$ by decomposing it into components parallel and orthogonal to $\widehat{\bv}_E$.  By dotting with $(\bI - \widehat{\bv}_E \widehat{\bv}_E)$ and $\widehat{\bv}_E$ respectively then solving for the relevant component of $\bG_\perp$, we find
\begin{equation}
\begin{split}
	\left( \bI - \widehat{\bv}_E \widehat{\bv}_E \right) \cdot \bG_\perp &= \frac{ \left( \bI - \widehat{\bv}_E \widehat{\bv}_E \right) \cdot \bF_{BFV,\perp}}{1 - \eta^2 / 2}, \\
	\widehat{\bv}_E \cdot \bG_\perp &= \frac{2}{\eta^2} \left( \widehat{\bv}_E \cdot \bF_{BFV,\perp} + F_{BFV,\parallel} \frac{v_\parallel}{v_\perp} \right).
\end{split}
\end{equation}
From here, it is straightforward to write an expression for $\bG_\perp$ in the case $u < v_E$:
\begin{equation} \label{eq:Gperpvedom}
	\bG_\perp = \frac{ \left( \bI - \widehat{\bv}_E \widehat{\bv}_E \right) \cdot \bF_{BFV,\perp}}{1 - \eta^2 / 2} + \frac{2}{\eta^2} \widehat{\bv}_E \left( \widehat{\bv}_E \cdot \bF_{BFV,\perp} + F_{BFV,\parallel} \frac{v_\parallel}{v_\perp} \right).
\end{equation}

We can now see by comparing \eqref{eq:Gperpvedom} to \eqref{eq:Gperpudom} - along with the knowledge that $G_\parallel = F_{BFV,\parallel}$ - that the following is a uniformly valid expression for $\bG$:
\begin{equation} \label{eq:genGdef}
	\bG = F_{BFV,\parallel} \bb + \frac{ \left( \bI - \widehat{\bv}_E \widehat{\bv}_E \right) \cdot \bF_{BFV,\perp}}{1 - \eta^2 / 2} + \frac{2}{\eta^2} \widehat{\bv}_E \left( \widehat{\bv}_E \cdot \bF_{BFV,\perp} + \mathbbm{1}_{v_E > u} F_{BFV,\parallel} \frac{v_\parallel}{v_\perp} \right),
\end{equation}
where $\mathbbm{1}_A$ denotes the indicator function, i.e., $1$ if $A$ holds and zero otherwise.  

For completeness, we summarize here the full form of our proposed energy-conserving, asymptotic preserving particle update:
\begin{equation} \label{eq:schemesummary}
\begin{split}
	\bv^{n+1} &= \bv^n + \dt \frac{q}{m} \left( \bE + \bv \times \bB \right)\nph \\
		&\hspace{1em} + \frac{\dt}{m} \left\{ \left[\bb - \frac{v_\parallel}{v_\perp} \frac{\bv_\perp}{v_\perp} \right] G_\parallel + \left[ \bI - \frac{\bv_\perp \bv_\perp}{v_\perp^2} \right] \cdot \bG_\perp \right\}^{n+1/2}, \\
		\bx^{n+1} &= \bx^n + \dt \bv\nph,
\end{split}
\end{equation}
where $\bG$ is given by \eqref{eq:genGdef} with all velocities evaluated at $t\nph$, and $\bF_{BFV}$ is defined by \eqref{eq:FBFVdef}.

There are three interesting points that warrant elaboration here.  Firstly, the relative size of $u$ and $v_E$ is dependent on $\dt$.  This is because the relevant values of $u$ and $v_E$ are those at $t\nph$; while $\bv_E$ changes little in a time-step, the rapid variation of $\bu$ has already been shown to lead it to shrink for increading $\Oc \dt$ - see \eqref{eq:vnphsmall}, where we have assumed $\bE = 0$ so $\bv_\perp \approx \bu$.  The consequences of this fact will be expanded upon in the following sections.  

Secondly, note that $\bG_\perp$ diverges as $u/v_E \rightarrow 0$ (i.e.\ $\Oc \dt \rightarrow \infty$ if $v_E \neq 0$).  This may initially appear alarming, but is accounted for by the fact that $\bG_\perp$ itself is never actually used in the velocity update - only its projection onto the orthogonal complement of $\bv_\perp$.  In the same limit $u/v_E \rightarrow 0$, $\bv_E$ dominates $\bv_\perp$, so that this projection is very nearly onto the orthogonal complement of $\bv_E$.  Since the divergence of $\bG_\perp$ is in the $\bv_E$ direction, this projection results in the actual effective force remaining well-behaved as $\eta \rightarrow 0$.  The reliance of the scheme on this precise cancellation, though, does indicate that it is crucial that the implicit solve be performed accurately.  In particular, $\eta$ should be computed self-consistently (as opposed to using some explicit estimation of $\eta$) since the scheme is extremely sensitive to its precise value when $\eta \ll 1$.  

Thirdly, one can see that $\bG$ is in general discontinuous on the manifold defined by $u = v_E$.  Indeed, 
\begin{equation}
	\lim_{u \rightarrow v_E^-} \bG - \lim_{u \rightarrow v_E^+} \bG = 2 \widehat{\bv}_E F_{BFV,\parallel} \frac{v_\parallel}{v_E}
\end{equation}
due to the presence of the indicator function.  This, again, is a consequence of the qualitative difference in the behavior of $\bv_\perp / v_\perp$ over a gyro-orbit depending on which of $u$, $v_E$ is larger. Of course, by construction, the gyroaverage of the effective force arising from $\bG$ is continuous, so this discontinuity is only expected to be an issue for the small subset of particles for which $u$ and $v_E$ are very close to equal, so that there is a risk of their relative size changing from time-step to time-step. Our adaptive time-stepping strategy outlined below will endeavor to choose time-steps so that this is not the case.

\subsection{Time-step Restrictions}
While an asymptotic preserving scheme admits the use of time-steps with $\Oc \dt \gg 1$, one of course cannot expect accurate results for arbitrarily large time-steps.  Understanding the restrictions on $\Delta t$ is crucial when applying the scheme to problems of scientific interest, as failure to do so risks viewing erroneous results as legitimate.  

An obvious restriction is that the time-step should be sufficiently small that variations in $\bE$ and $\bB$ are well-resolved.  The scheme proposed here features two additional time-step restrictions that we now describe. After deriving the time-step restrictions on the scheme, we present an adaptive time-stepping scheme designed to ensure these constraints are respected at every step.

Like the MI scheme but unlike BFV, our proposed scheme \eqref{eq:schemesummary} has an effective force that only approximates $-\mu \nabla B$ in a gyroaveraged sense.  That is, we must rely on the time-stepping process to perform a sort of ``implicit" gyroaverage for us - here we use the term ``implicit" not in the sense that a system of equations must be solved, but in the sense that no gyroaverage operator explicitly appears in our time integration scheme.  This fact was alluded to when discussing MI; we discuss it in considerably more detail here.  

Recall that the $\nabla B$ drift in the perpendicular direction that we seek to capture has the form
\begin{equation}
	\bv_{\nabla B} = \frac{\bb}{\Oc} \times \frac{\mu}{m} \nabla B.
\end{equation}
The scheme proposed here, on the other hand, gives rise to a drift of the form
\begin{equation}
	\bv_{\nabla B}^\textrm{AP} = \frac{\bb}{\Oc} \times \frac{1}{m} \left( \mu_\textrm{eff}^\textrm{CN} \nabla B - \bF_{cons} \right).
\end{equation}
The scheme thus features an anomalous drift with velocity 
\begin{equation}
	\Delta \bv_{\nabla B} = \bv_{\nabla B} - \bv_{\nabla B}^\textrm{AP}.
\end{equation}

By construction, the gyroaverage of this anomalous displacement vanishes.  However, on shorter time-scales, it has the potential to impact the particle's long-term trajectory.  This can be avoided by enforcing two conditions: (a) the anomalous displacement due to this anomalous drift should be no larger than the gyroradius, and (b) the time-scale over which this anomalous displacement averages to zero should be small compared to time-scales of interest, denoted henceforth by $\tau_\textrm{res}$.  

Deriving the restrictions on time-step arising from these conditions is quite challenging in general, but can be done analytically in two important limits: (i) $u\nph \gg v_E$, and (ii) $v_E \gg u\nph$.  The derivations of these restrictions are lengthy, and therefore confined to Appendix B.  We summarize the results in Table 1, in which we record the maximum value of $\Oc \dt$ permitted by each restriction in each of the two asymptotic limits above.
\begin{table}[t]
\centering
\vspace{1em}
\caption{The maximum allowable value of $\Oc \dt$ to ensure (a) the anomalous displacement does not exceed gyroradius and (b) the time-scale over which accurate implicit gyroaveraging is achieved is shorter than $\tau_{\textrm{res}}$.  Expressions are valid in the asymptotic limits (i) $u\nph \gg v_E$ and (ii) $u\nph \ll v_E$, respectively.} \label{table:dt_table}
\begin{tabular}{c | c c}
 $\max \left( \Oc \dt \right)$ & (a) Displacement restriction & (b) Avg. restriction \\
 \hline
 (i) $u^{n+1/2} \gg v_E$ & $2 \min \left\{ \sqrt{2} (\delta_\perp)^{-1/2},  (\delta_\parallel)^{-1} \right\}$ & $2 \left( \frac{\Omega_c \tau_{\textrm{res}}}{\pi} \right)^{1/2}$ \\
 (ii) $u^{n+1/2} \ll v_E$ & $\sqrt{2} \left( \delta_E + \delta_\parallel \right)^{-1/2}$ & $\Omega_c \tau_{\textrm{res}}$
\end{tabular}
\vspace{1em}
\end{table}

The relevant non-dimensional quantities in determining these restrictions on $\Oc \dt$ are defined by
\begin{equation} \label{eq:deltadefs}
	\delta_\perp = \rho \frac{\left\| (\nabla B)_\perp \right\|}{B}, \qquad \delta_\parallel = \frac{v_\parallel}{\Oc} \frac{\left\| (\nabla B)_\parallel \right\|}{B}, \qquad \delta_E = \frac{v_E}{\Oc} \frac{\left\| (\nabla B)_\perp \right\|}{B}.
\end{equation}
The quantity $\tau_{\textrm{res}}$ denotes the shortest time-scale in our problem that we wish to resolve.  For brevity, we quote here only simplified restrictions under the assumptions $\delta_\parallel, (\Oc \tau_\textrm{res})^{-1} \ll 1$.  The more general expressions may be found in Appendix B.  

\subsection{Adaptive Time-stepping}
Having understood the limitations on our scheme's time-step, we propose an adaptive time-stepping strategy that respects these limits.  In addition, the time-stepping strategy should endeavor to avoid the $u\nph = v_E$ manifold on which our effective force is discontinuous.  We specify parameters $\alpha$, $\beta$, $\epsilon \in (0,1)$ and $\Gamma > 0$. We then select a time-step as follows:
\begin{enumerate}
	\item Compute all relevant quantities - $\bE$ and $\bB$, their gradients, $\bv_E$, etc.\ - at the current particle position.
	\item Set $(\Oc \dt)_E = \alpha \min \left\{ \sqrt{2} \left( \delta_E + \delta_\parallel \right)^{-1/2}, \Oc \tau_{res}, \Gamma \delta_\perp^{-1} \right\}$.
	\item Estimate $\bu$ at the half time-step with $\tilde{u}^{n+1/2} = \left\| \bv_\perp - \bv_E \right\|/\sqrt{1 + (\Oc \dt)_E^2/4}$.
	\item If $v_E > (1 + \beta) \tilde{u}^{n+1/2}$, set $\dt = (\Oc \dt)_E / \Oc$ and proceed with the time-step.
	\item Else, if $v_E > (1 - \beta) u^n$, set $\dt = \min \{ \tau_{\textrm{res}}, \Gamma \Oc^{-1} \}$ and proceed with the time-step.
	\item Else, set $(\Oc \dt)_u = \alpha \min \left\{ 2 \min \{ \sqrt{2}/\sqrt{\delta_\perp}, 1/\delta_\parallel \}, 2 \sqrt{ \Oc \tau_{\textrm{res}}/\pi} \right\}$.  Then set $\dt = (\Oc \dt)_u / \Oc$ and proceed with the time-step.
	\item If, upon completion of the time-step, it is found that the fractional change in magnetic moment $\Delta \mu / \mu$ during the time-step exceeds $\epsilon$, shrink the step size by a factor of $\alpha \epsilon \mu / \Delta \mu$ and recompute the step.  
\end{enumerate}

Several details of this time-step selection warrant elaboration.  First, the parameters $\alpha$, $\beta$ and $\Gamma$ have simple interpretations.  $\alpha$ controls how close to the time-step restrictions of Table \ref{table:dt_table} one is willing to get.  $\alpha = 1$ corresponds to steps at the limit of what is valid, while smaller values of $\alpha$ correspond to more conservative strategies.  In our numerical experiments, we use $\alpha = 0.9$.  

The parameter $\beta$ controls how close to the $u\nph = v_E$ manifold - on which our effective force is discontinuous - we allow ourselves to approach.  If there is no way to respect the time-step constraints while keeping either $v_E / u\nph > 1+\beta$ or $v_E / u\nph < 1-\beta$, then the gyroperiod is resolved (see step 5) so that the discontinuity is no longer a concern.  In our numerical experiments, we use $\beta = 0.2$.  Note that the circumstances in which this restriction is active are extremely rare.  Indeed, being required to resolve $\Oc^{-1}$ requires both that $v_E / u^n \in (1-\beta, 1+\beta)$ \textit{and} the maximum allowable time-step to be not much larger than $\Oc^{-1}$.  In quasi-neutral plasmas, one typically expects $v_E \ll u^n$, so this constraint only applies to a select few particles for which a large time-step was impossible in the first place.

The parameter $\Gamma$ measures the accuracy with which we wish to resolve spatial variations in the magnetic field.  Indeed, note that one component of the constraint in step 2 reads
\begin{equation}
	\Delta t \leq \Gamma \left( u \frac{ \left\| (\nabla B)_\perp \right\|}{B} \right)^{-1}.
\end{equation}
The right-side measures fractional variation in the magnetic field due to perpendicular velocity.  $\Gamma = 1$ corresponds to resolving the spatial scale over which $B$ doubles.  In our numerical experiments, we choose $\Gamma = 0.1$, corresponding to resolving $10\%$ variations in $B$.  Note that $\Gamma$ also appears in the time-step chosen when $\Oc^{-1}$ must be resolved (step 5).  It is not necessary that these two constants be equal, but $0.1$ turns out to be a reasonable value for each.  

The parameter $\epsilon$ controls the maximum permissible fractional change in $\mu$ within a time-step.  It appears in step 7 as a safeguard for situations in which the particle is weakly magnetized.  When $\mu$ varies non-adiabatically, it is typically important to resolve these changes in $\mu$.  This is ensured by choosing a relatively small value of $\epsilon$.  Our numerical tests use $\epsilon = 0.15$.  

A final note regarding $\tau_{\textrm{res}}$: In general, this is a problem-dependent quantity to be determined by the user.  A reasonable starting point, however, that is used in the majority of our numerical tests is as follows.  Begin by defining lengths scales
\begin{equation}
	L_\perp = \Gamma \left( \frac{ \left\| (\nabla B)_\perp \right\|}{B} \right)^{-1}, \qquad L_\parallel = \Gamma \left( \frac{ \left\| (\nabla B)_\parallel \right\|}{B} \right)^{-1}, \qquad L_E = \Gamma \left( \frac{\nabla v_E}{v_E} \right)^{-1},
\end{equation}
arising from perpendicular and parallel variations in $B$ and $v_E$.  A time-scale to be resolved can then be defined by
\begin{equation}
	\tau_{res} = \min \left\{ \frac{L_\perp}{v_E}, \frac{L_\parallel}{v_\parallel}, \frac{L_E}{v_E} \right\}.
\end{equation}

\section{Numerical Examples}
We test our scheme by investigating single-particle motion in four electromagnetic field configurations.  We use a simple magnetic mirror configuration, a case with transverse magnetic field gradient in a uniform electric field, a case in which the particle passes through a weakly magnetized region, and a Solov'ev equilibrium in tokamak geometry with a flux-function electric field.  

In all cases, we work in a dimensionless formulation in which $q/m = 1$, and measure both accuracy and conservation properties of the scheme.  To measure accuracy, we compare to a simulation using the standard Boris integrator - used by the majority of PIC schemes - with $\Oc \dt \leq 0.1$.  We solve the systems of equations arising from our implicit discretizations using a Jacobian-Free Newton Krylov (JFNK) method with tolerance set to $10^{-12}$.  

The adaptive time-stepping scheme and definition of $\tau_{\textrm{res}}$ found above are used except where otherwise noted.   When individual values of $\Oc \dt$ are reported, these are averages over the entirety of the simulation.  When noteworthy, we plot the value of $\Oc \dt$ as a function of time to illuminate the behavior of the adaptive time-stepping scheme.  

In each test case, we compare our scheme against not only a well-resolved Boris run, but also the BFV scheme, MI, and unmodified Crank-Nicolson.  Each scheme uses the adaptive time-stepping scheme described above for a fair comparison.  

\subsection{Magnetic Mirror}
As is standard, our magnetic mirror is formed by two circular current loops.  Let the loops be oriented normal to the $z$-axis, centered at $(0,0,\pm L/2)$, with radius $r$ and carrying current $I$.  The resulting magnetic field (up to constants) near the $z$-axis is given by
\begin{equation}
	\bB = \frac{Ir^2}{2} \sum_{\pm = +, -} \left( \frac{1}{\left[ r^2 + \left( z \pm L/2 \right)^2 \right]^{3/2} } \right) \left( \begin{array}{c} \frac{3x}{2} \left( \frac{ z \pm L/2 }{ r^2 + \left( z \pm L/2 \right)^2 } \right) \\ \frac{3y}{2} \left( \frac{ z \pm L/2 }{ r^2 + \left( z \pm L/2 \right)^2 } \right) \\ 1 \end{array} \right).
\end{equation}

We work with $r = 3$, $L = 15$, $I = 3 \times 10^4$, leading to a mirror ratio of $R_{mirr} = 9.8342$.  We initialize the particle at $\bx^0 = 0.02 \widehat{\mathbf{y}}$ with $\bv^0 = \widehat{\bx} + v^0_\parallel \widehat{\mathbf{z}}$.  To a good approximation, this results in gyration about the $z$-axis.  Having fixed the perpendicular velocity at $1$, the trapped-passing boundary may be expressed purely in terms of a critical initial parallel velocity $v_{crit}$ given by
\begin{equation}
	v_{crit} = \sqrt{ R_{mirr} - 1 }.
\end{equation}

\begin{figure}[h]
	\centering
	\includegraphics[width=0.49\textwidth]{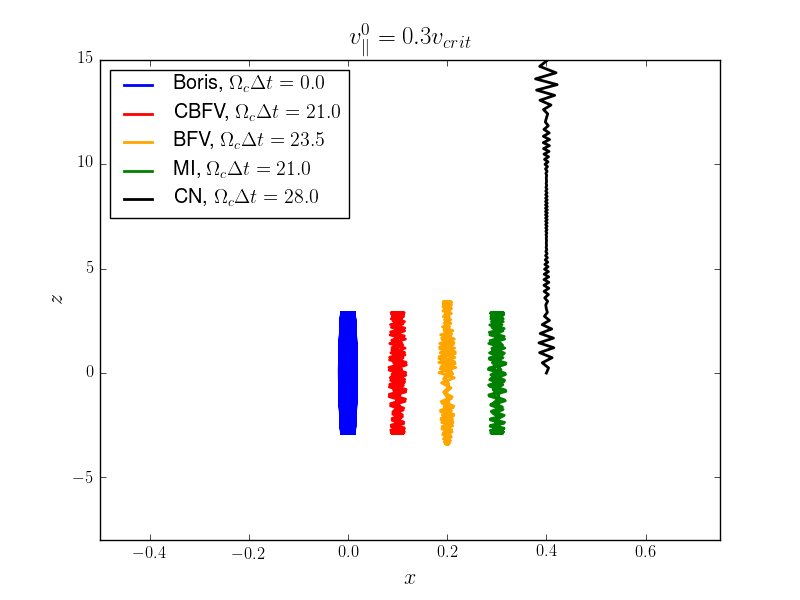}
	\includegraphics[width=0.49\textwidth]{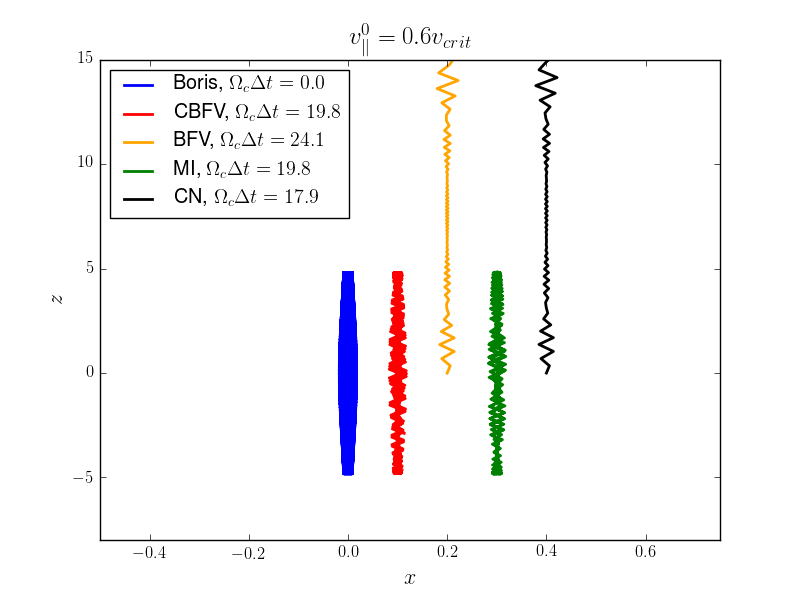}
	\includegraphics[width=0.49\textwidth]{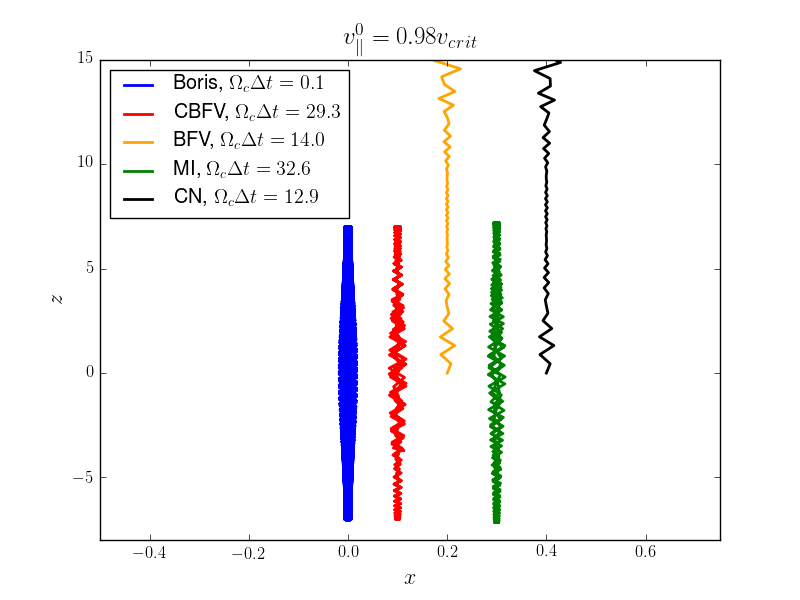}
	\includegraphics[width=0.49\textwidth]{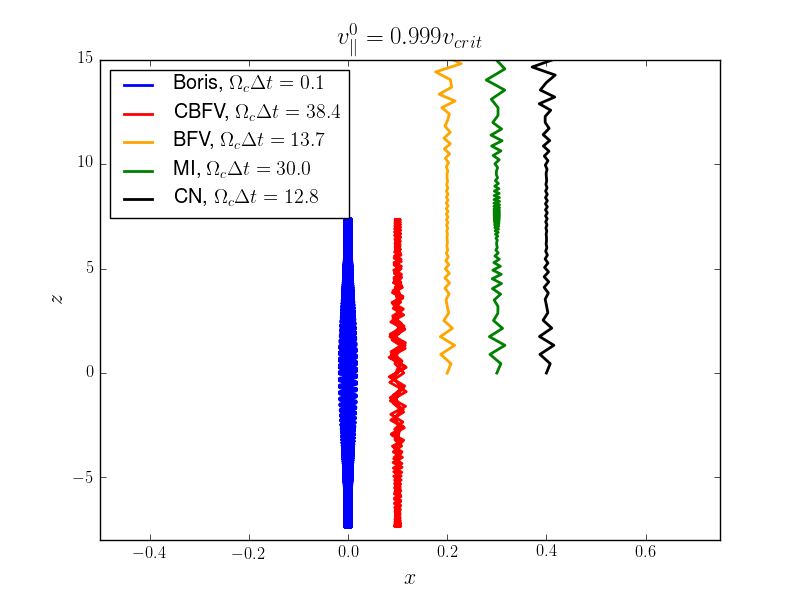}
	\caption{Trajectories in the $x$-$z$ plane for the magnetic mirror test case.  Schemes tested are fully-resolved Boris (blue), our new scheme denoted CBFV (red), BFV (orange), MI (green), and Crank-Nicolson (black).  Trajectories are offset by 0.1 in the $x$-direction for visualization purposes.  The data plotted for initial parallel velocity of $0.3v_{crit}$ (top left), $0.6v_{crit}$ (top right), $0.98v_{crit}$ (bottom left), and $0.999v_{crit}$ (bottom right).} \label{fig:mirrorxz}
\end{figure}

We perform a parameter scan in the initial parallel velocity $v_\parallel^0$ to test how accurately each scheme captures the trapped-passing boundary.  The particle trajectories in the $x$-$z$ plane for several particular values of $v_\parallel^0$ are found in Figure \ref{fig:mirrorxz}.
As expected, Crank-Nicolson dramatically underestimates the mirror force for $\Oc \dt \gg 1$, and confinement is not achieved for any of the parallel velocities tested.  BFV correctly predicts confinement for parallel velocities relatively small compared to $v_{crit}$, but underestimates the trapped-passing boundary by more that $40\%$.  

\begin{figure}[h]
	\centering
	\includegraphics[width=0.6\textwidth]{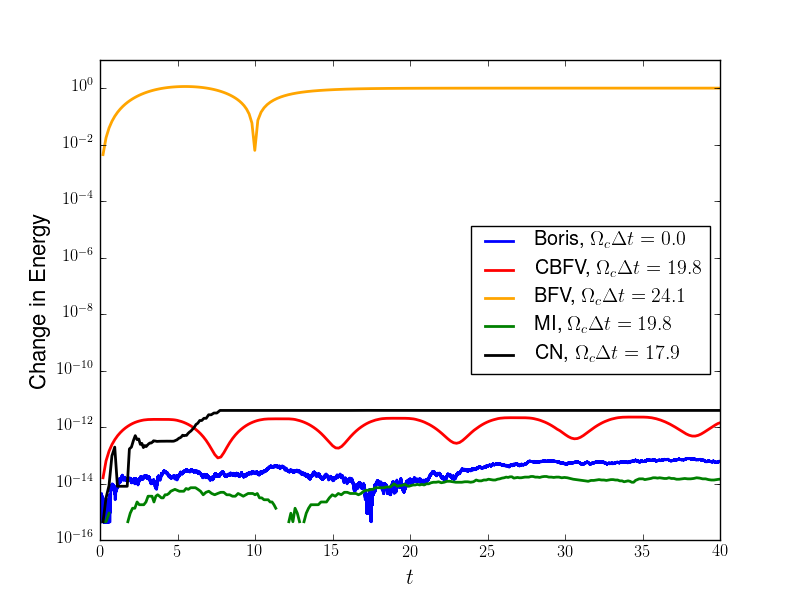}
	\caption{Change in particle energy during a simulation for magnetic mirror configuration.  The plotted trajectories have $v_\parallel^0 = 0.6v_{crit}$, but are representative of the behavior for all tested values of $v_\parallel^0$.}
	\label{fig:econsmirror}
\end{figure}

In this case, both the present scheme and MI perform quite well, predicting the trapped-passing boundary quite accurately.  An understanding of the reason for this can be gleaned from Figure \ref{fig:econsmirror}, in which we measure energy conservation during a sample simulation.
As expected, our scheme and Crank-Nicolson conserve energy up to the tolerance of the nonlinear solver.  In the absence of an electric field, MI and Boris conserve energy up to numerical round-off.  BFV, on the other hand, features $O(1)$ errors in energy due to the form of the effective force introduced, leading to its poor performance.  

\begin{figure}[h]
	\centering
	\includegraphics[width=0.49\textwidth]{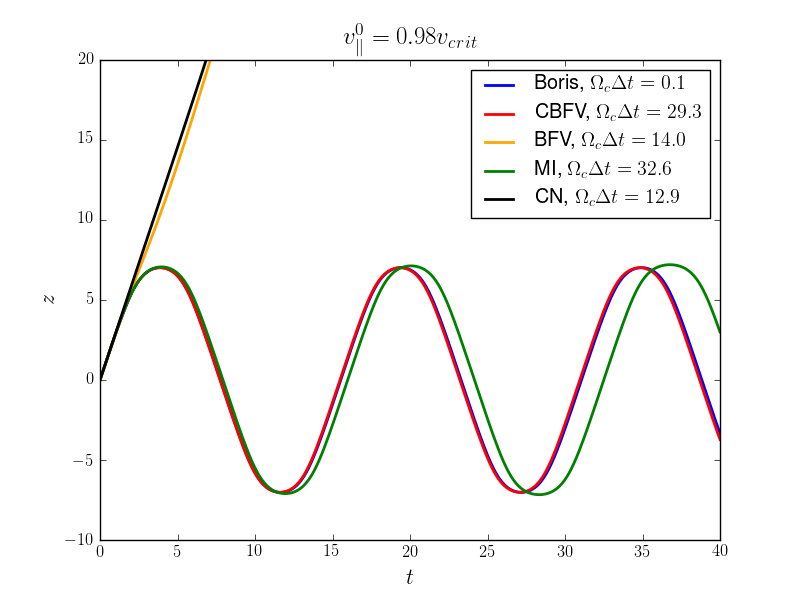}
	\includegraphics[width=0.49\textwidth]{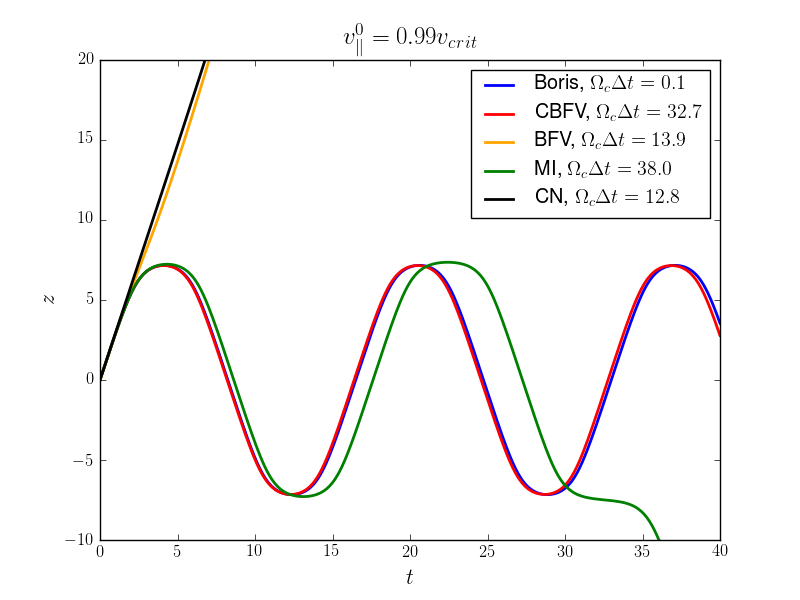}
	\includegraphics[width=0.49\textwidth]{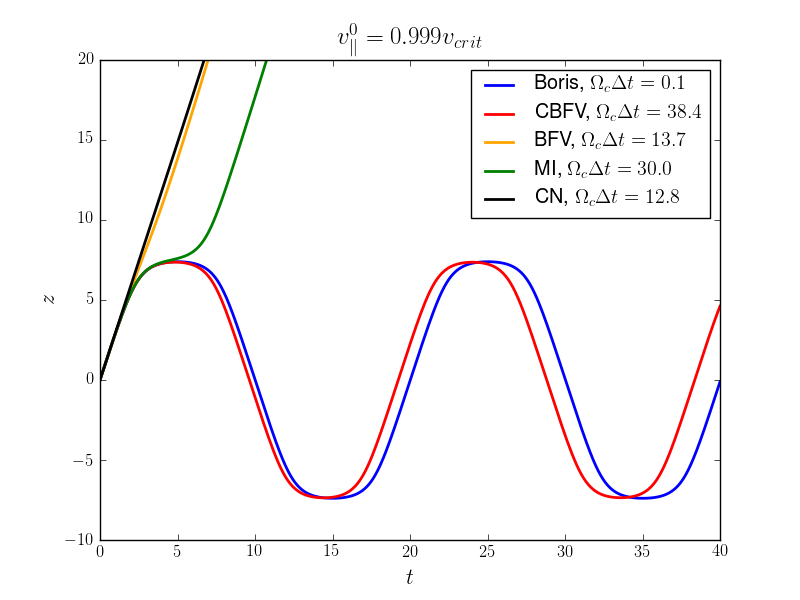}
	\includegraphics[width=0.49\textwidth]{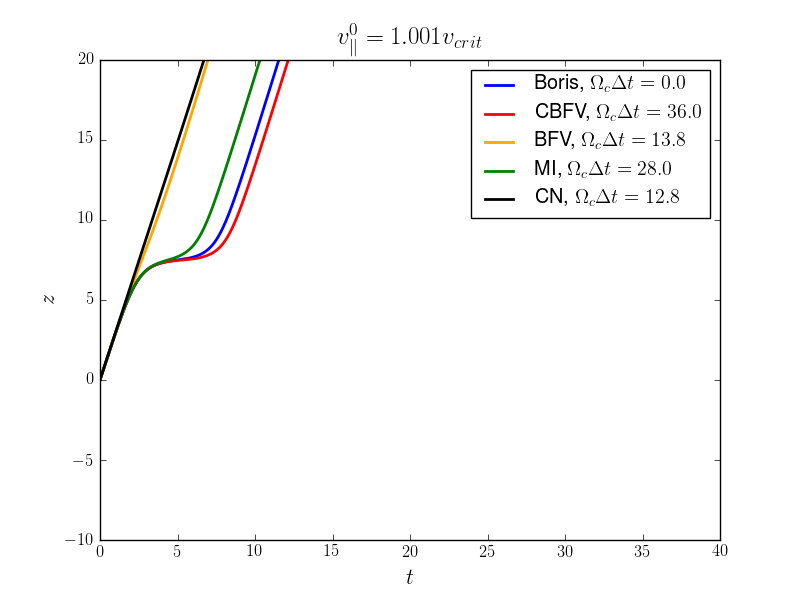}
	\caption{The $z$ coordinate as a function of $t$ for the mirror test.  We focus on initial parallel velocities near the trapped-passing boundary at $v_{crit}$.  We use $0.98v_{crit}$ (top left), $0.99v_{crit}$ (top right), $0.999v_{crit}$ (bottom left), and $1.001v_{crit}$ (bottom right) to demonstrate that the new scheme captures the bounce frequency well and the trapped-passing boundary to within $0.01\%$.} \label{fig:zoftmirror}
\end{figure}

In addition to the $x$-$z$ plane, we plot $z$ as a function of time for each method in Figure \ref{fig:zoftmirror}, this time focusing on values of $v_\parallel^0$ near $v_{crit}$.
When focusing on values of $v_\parallel^0$ near the trapped-passing boundary, the improved performance of the new scheme, even relative to MI, is readily visible.  

\subsection{Transverse Magnetic Field Gradient}
The previous test exercises only the mirror force, in the absence of an electric field.  Here, we test each scheme's ability to capture the transverse $\nabla B$ drift in a homogeneous electric field.  We set $\bB = (10 + x) \widehat{\mathbf{z}}$, $\bE = \widehat{\mathbf{y}}$, and initialize the particle at the origin with $\bv^0 = \widehat{\mathbf{x}}$.  Simulations are run to the final time $T = 300$.  

\begin{figure}[h]
	\centering
	\includegraphics[width=0.49\textwidth]{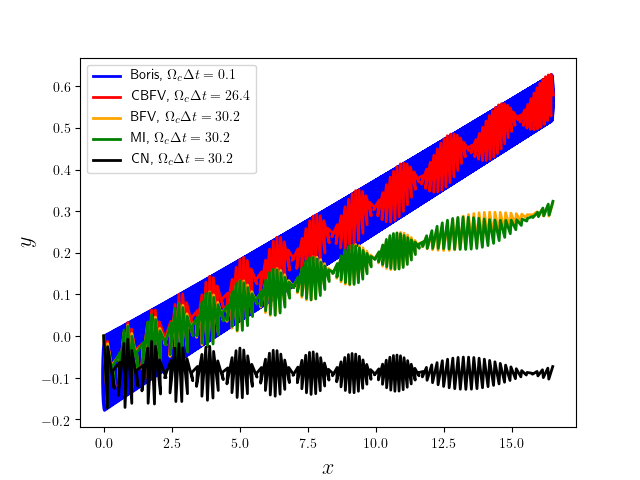}
	\includegraphics[width=0.49\textwidth]{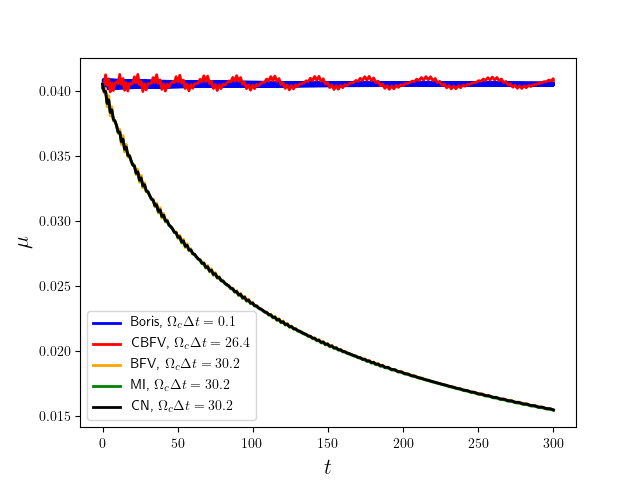}
	\caption{The trajectory in the $x$-$y$ plane (left) and time history of magnetic moment $\mu$ (right) for the second test case with transverse magnetic field gradient in uniform electric field.}
	\label{fig:ExBandGradB1}
\end{figure}

These fields induce an $\bE \times \bB$ drift in the positive $x$-direction and a $\nabla B$ drift in the positive $y$-direction.  This test case affords the opportunity to measure magnetic moment conservation and ability to capture the polarization drift, the latter arising because the $\bE \times \bB$ speed is time-varying.  The trajectory in the $x$-$y$ plane as well as $\mu$ as a function of time appear in Figure \ref{fig:ExBandGradB1}.

\begin{figure}[h]
	\centering
	\includegraphics[width=0.49\textwidth]{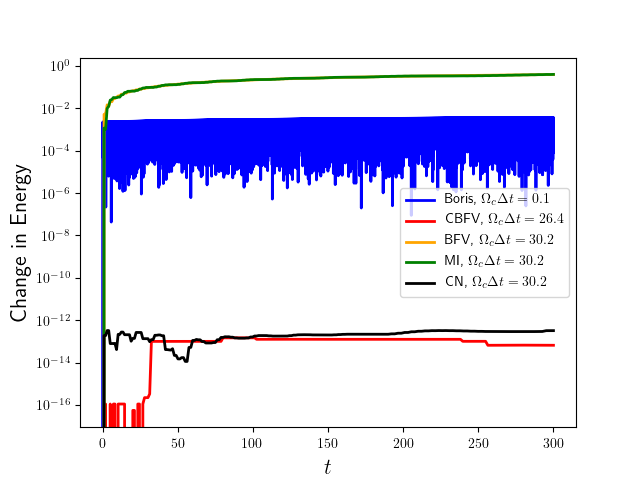}
	\includegraphics[width=0.49\textwidth]{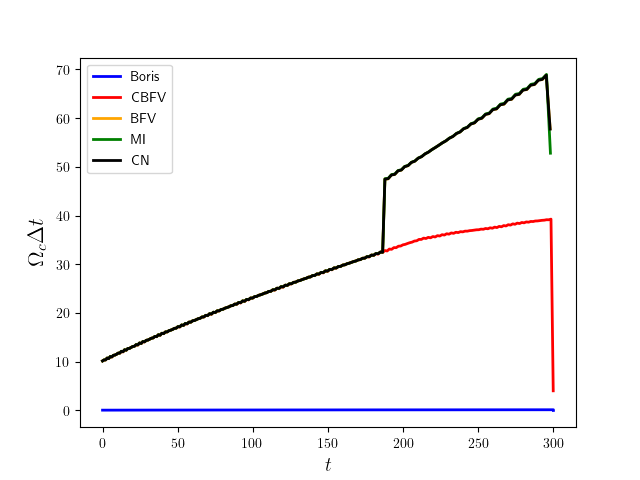}
	\caption{Energy errors (left) and time history of $\Oc \dt$ selected by our adaptive time-stepping strategy (right) for second test case with transverse magnetic field gradient in fixed electric field.  Note that the distinction in time-step size arises from the shrinking of the gyration velocity for the non-conservative schemes.}
	\label{fig:ExBecons}
\end{figure}

As is readily apparent, the new scheme features improved accuracy in capturing the trajectory.  This can be directly linked to improved magnetic moment conservation: other schemes underestimate $\mu$, so they also underestimate the $\nabla B$ drift, for it is proportional to $\mu$.  The reason for the failure of earlier schemes - especially BFV - to conserve $\mu$ is simple to understand.  Fundamentally, adiabatic invariance of $\mu$ is a result of the energy equation
\begin{equation}
	\frac{m}{2} \deriv{v^2}{t} = q \bv \cdot \bE,
\end{equation}
followed by substituting in the drift motion for $\bv$, gyroaveraging, and neglecting small terms.  When the ``effective" force $-\tilde{\mu} \nabla B$ is introduced as in BFV, a new term appears in the energy equation above:
\begin{equation}
	\frac{m}{2} \deriv{v^2}{t} = q \bv \cdot \bE - \tilde{\mu} \bv \cdot \nabla B.
\end{equation}
In addition to leading to $O(1)$ reduction in total energy - see Figure \ref{fig:ExBecons} - this term also means that when retracing the steps in the $\mu$-conservation derivation, one must find $\dot{\mu} \approx -\tilde{\mu} \bv_E \cdot \nabla B$, where we have again assumed that $\bv_E$ dominates the drift motion.  In our case, $\bv_E$ and $\nabla B$ are parallel, so for $\Oc \dt \gg 1$ (i.e. $\mu \approx \tilde{\mu}$), one expects exponential decay of $\mu$ at rate $v_E \left\| \nabla B \right\|$, just as observed.  

Crank-Nicolson, on the other hand, does conserve energy.  However, its underestimation of the $\nabla B$ drift keeps the particle in a region of artificially high electrostatic potential.  This must be compensated for by robbing the particle of kinetic energy.  Since we've already shown that the $\bE \times \bB$ drift is accurately captured, that kinetic energy must come from the particle's gyration, thus leading to decay of $\mu$.  

Note also that while MI conserves energy to numerical roundoff in the absence of an electric field - see magnetic mirror test case above - it features $O(1)$ errors in the presence of even a uniform electric field due to its usage of an effective velocity.  Finally, we note that our adaptive time-stepping strategy chooses larger time-steps for BFV, MI, and Crank-Nicolson than for our scheme near the end of the simulation.  This is a result of the shrinking $\mu$ value those schemes experience, which shrinks the gyroradius and leads the scheme to believe a larger time-step is possible.

\subsection{Weakly Magnetized Region}
A strong motivation for an AP scheme capable of stepping over the gyrofrequency is that, unlike drift- or gyro-kinetics, it can handle weakly magnetized regions in which the guiding-center approximation is invalid.  We test this capability in a simple geometry by setting
\begin{equation}
	\bB = \widehat{\mathbf{z}} \sqrt{ 1 + 4(x - 12)^2 }, \qquad \bE = \widehat{\mathbf{y}}.
\end{equation}
Having already convinced ourselves that C-N fails to capture the $\nabla B$ drift for $\Oc \dt \gg 1$, we omit it in plots here to reduce clutter.

\begin{figure}
	\centering
	\includegraphics[width=0.49\textwidth]{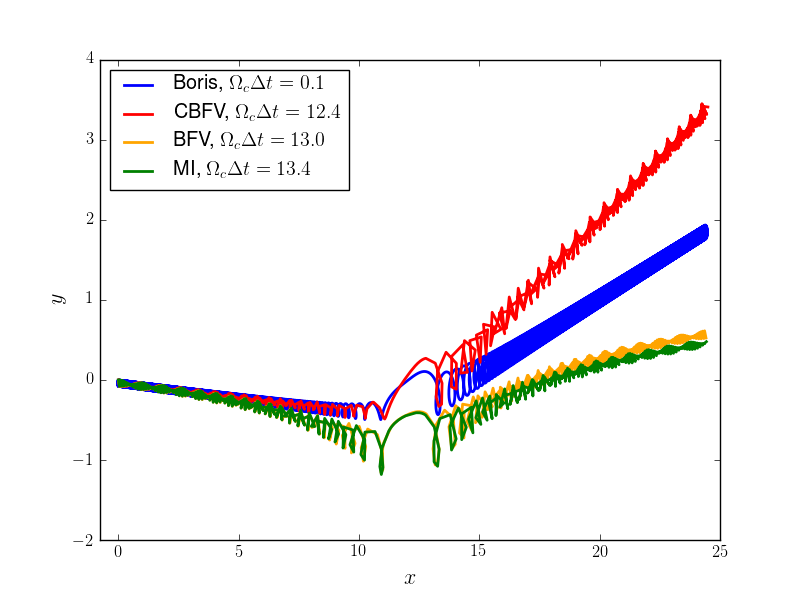}
	\includegraphics[width=0.49\textwidth]{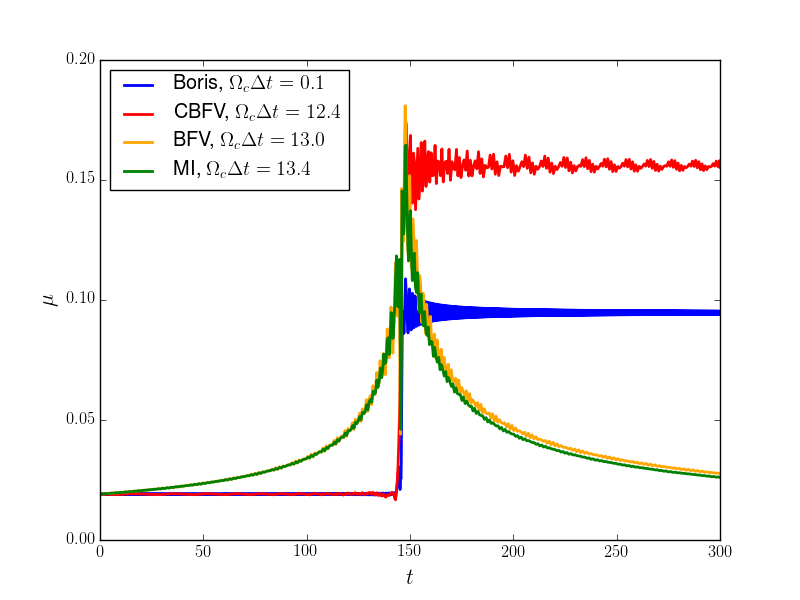}
	\caption{Particle trajectories in the $x$-$y$ plane (left) and time history of magnetic moment $\mu$ for the third test case featuring a weakly magnetized region.  Note that the size of the rapid, non-adiabatic jump in $\mu$ depends sensitively on the gyrophase of the particle as it approaches $x = 12$.  Thus, no scheme that steps over the gyroperiod can accurately reproduce it for any given particle.  However, we observe again that the present scheme is the only one that conserves $\mu$ in the strongly magnetized regions while stepping over gyroperiod.}
	\label{fig:NonAdiabatTraj}
\end{figure}

We initialize a particle at the origin, where it experiences an $\bE \times \bB$ drift in the positive $x$-direction.  This pushes it through the narrow region surrounding $x=12$ in which the particle is quite weakly magnetized and the guiding center approximation is violated.  Therein, it experiences a large, non-adabatic jump in magnetic moment $\mu$ before continuing to drift to the right into another strongly magnetized region.  A sample trajectory and time history of $\mu$ for each scheme appear in Figure \ref{fig:NonAdiabatTraj}.

Note that, while our proposed scheme is the only one able to reproduce the adiabatic invariance of $\mu$ in the magnetized regions far from $x = 12$, it still does not recover the particle trajectory accurately.  This is because the rapid, non-adiabatic jump in $\mu$ that occurs as the particle passes through the weakly magnetized region is \textit{very} sensitive to the particle's gyrophase as it enters the weakly magnetized region.  Since stepping over the gyroperiod inherently throws away gyrophase information, no scheme using $\Oc \dt \gg 1$ can hope to recover the details of any given particle trajectory in this case.  

\begin{figure}[h]
	\centering
	\includegraphics[width=0.7\textwidth]{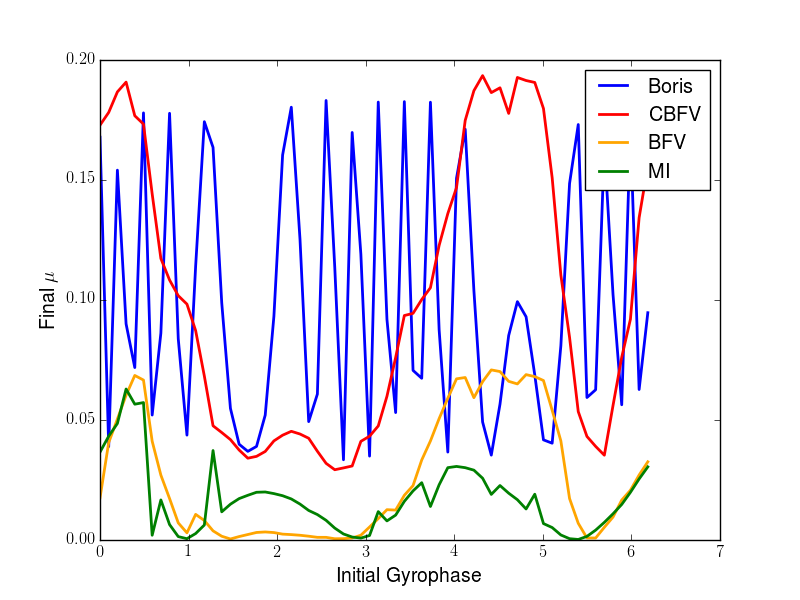}
	\caption{Final value of magnetic moment $\mu$ for various initial gyrophases in the third test case featuring a weakly magnetized region.}
	\label{fig:gyrophase}
\end{figure}

What one can hope for, however, is that the \textit{average} jump in magnetic moment is reproduced accurately when averaged over initial particle gyrophase.  Indeed, a fundamental assumption in the drift- and gyro-kinetic approximations is that the distribution function is independent of gyrophase.  In this case, recovering the average $\mu$ jump is sufficient to predict the ensemble dynamics.  To this end, we initialize particles at the origin with velocities $\bv^0 = \left( \cos \gamma, \sin \gamma \right)$, $\gamma = 2\pi n/ 64$, $n = 0, ..., 63$ and measure the final magnetic moment of the particle averaged over the interval $t \in [100, 120]$.  The final magnetic moment as a function of initial gyrophase for each scheme is plotted in Figure \ref{fig:gyrophase}.

As expected, the magnetic moment is not accurately reproduced for most particular gyrophases.  However, the present scheme reproduces the mean $\mu$ of the fully-resolved Boris run to within $0.8 \%$, and the standard deviation to within $14\%$.  For detailed values, see Table \ref{table:meanstdmu}.
\begin{table}
\centering
\vspace{1em}
\caption{Mean and standard deviation in final $\mu$ value for the third test case.}
\label{table:meanstdmu}
\begin{tabular}{c | c c c c}
	 & Boris ($\Oc \dt \ll 1$) & CBFV & BFV & MI \\
	 \hline
	 Mean $\mu$ & 0.099487 & 0.100280 & 0.026340 & 0.0172801 \\
	 Std.\ dev.\ in $\mu$ & 0.051293 & 0.058710 & 0.025910 & 0.014327
\end{tabular}
\vspace{1em}
\end{table}
Recall that a drift- or gyro-kinetic approximation would predict \textit{no} change in $\mu$ in this (or any other) case.

\begin{figure}
	\centering
	\includegraphics[width=0.49\textwidth]{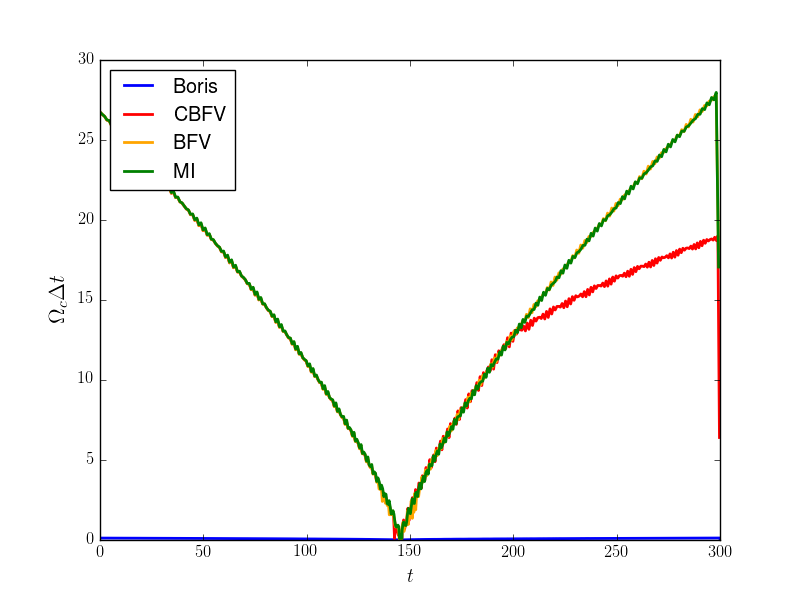}
	\includegraphics[width=0.49\textwidth]{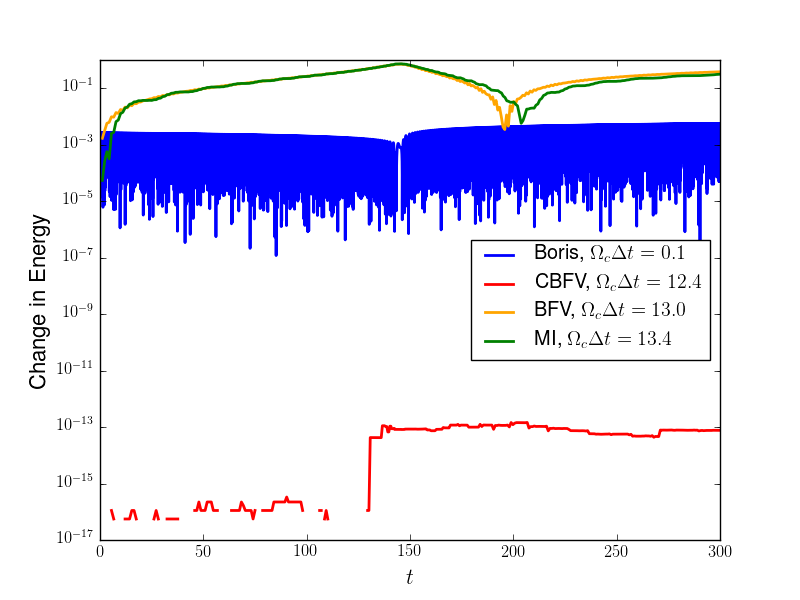}
	\caption{Non-dimensional time-step size $\Oc \dt$ as a function of time (left) and energy errors (right) for the third test case featuring a weakly magnetized region.  Missing data points in the energy plot for the BFV schemes correspond to times at which the change in energy is zero up to numerical round-off.}
	\label{fig:nonadiabatdt}
\end{figure}

As a final note, we plot $\Oc \dt$ as a function of time and the energy errors for this test case in Figure \ref{fig:nonadiabatdt}.  We use initial gyrophase $\gamma = 2\pi \times 63/64$ for these plots, as in Figure \ref{fig:NonAdiabatTraj}.  
As expected, our adaptive time-step strategy permits large time-steps except when the particle is in the weakly magnetized region, in which all scales are resolved.  Observe that, as in the previous test case, the underestimation of $\mu$ by BFV and MI leads them to take larger time-steps toward the end of the simulation.  

\subsection{Tokamak Equilibrium}
As a final test, we consider an equilibrium magnetic field in a tokamak-like geometry.  Simple analytic solutions of the Grad-Shafranov solutions described in \cite{cerfon2010one} specify the poloidal field, and the toroidal field is assumed to arise from a line current along the $z$-axis.  In particular, a Solov'ev equilibrium solution of the Grad-Shafranov equation for the flux function $\psi$ can be written as
\begin{equation}
	\psi (r, z) = \frac{C}{8} r^4 + d_1 + d_2 r^2 + d_3 \left( r^4 - 4 r^2 z^2 \right),
\end{equation}
where $C$, $d_i$ are constants and $(r, \phi, z)$ is the standard polar coordinate system.  The poloidal magnetic field is then given by $\bB_p = \nabla \psi \times \mathbf{e}_\phi / r$.  In \cite{pataki2013fast}, it is shown that the coefficients $d_i$ may be found uniquely by specifying the tokamak parameters $\varepsilon$ (inverse aspect ratio), $\kappa$ (elongation), and $\delta$ (triangularity).  Our poloidal field is thus uniquely specified by setting $C=300$ ($C$ simply controls the overall strength of the poloidal field) and using the shape parameters of the International Thermonuclear Energy Reactor (ITER): $\varepsilon = 0.32$, $\kappa = 1.7$, and $\delta = 0.33$.  The toroidal field is chosen to be $\bB_{tor} = 800 \mathbf{e}_\phi / r$.  

Additionally, we introduce an electrostatic potential, which is itself a flux function.  This choice is motivated by the fact that, in steady state, the ideal MHD Ohm's law ($\bE + \bv \times \bB = 0$) requires that $\bB \cdot \nabla \phi = 0$.  We define the electrostatic potential by $\phi = -\psi / 5$.  This non-trivial spatial dependence of the electric field necessitates an additional subtlety in order to retain energy conservation in the single-particle case.   Indeed, Crank-Nicolson only features exact energy conservation when the electric field varies \textit{linearly} in space, so that $\left( \bx^{n+1} - \bx^n \right) \cdot \bE\nph = \phi^n - \phi^{n+1}$ identically.  In \cite{simo1992exact}, a trick is introduced to recover exact energy conservation in the general case; in the present context, the trick consists of the replacement
\begin{equation}
	\bE\nph \rightarrow \frac{\phi^n - \phi^{n+1}}{\left( \bx^{n+1} - \bx^n \right) \cdot \bE\nph} \bE\nph
\end{equation}
in the velocity update equation.  Note that this replacement results in a scheme that is still second-order accurate, as the new pre-factor tends to one at second order.  Note further that this trick is not necessary for the total energy conservation theorems of implicit PIC - again, see \cite{chen2015multi, chen2011energy}.  We simply use it here demonstrate that exact conservation can be achieved in the single particle case as well.

\begin{figure}
	\centering
	\includegraphics[width=0.9\textwidth]{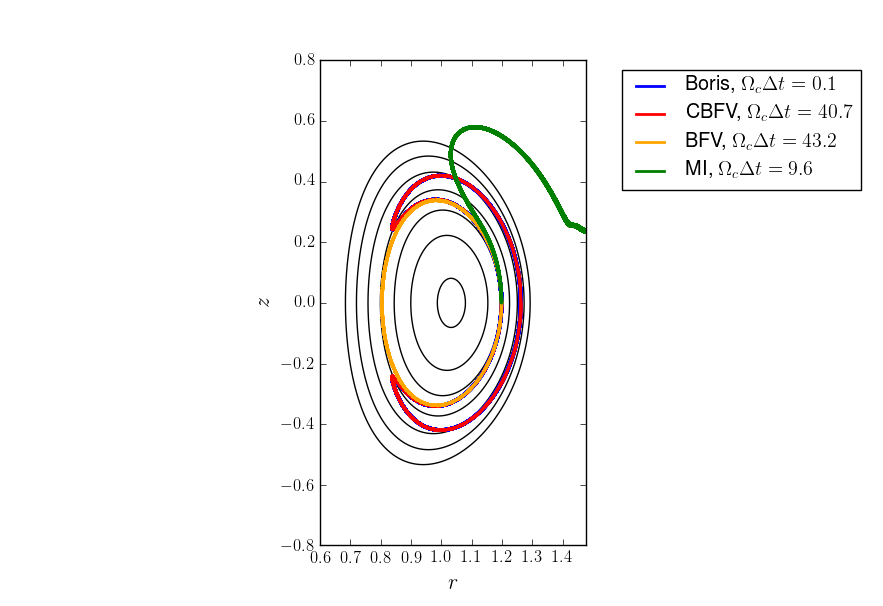}
	\caption{Banana orbits the $r$-$z$ plane in Solov'ev equilibrium magnetic field with flux-function electrostatic potential.  This figure uses $v_y^0 = 0.7$.}
	\label{fig:banana}
\end{figure}

We initialize a particle in these fields at $\bx = 1.2 \widehat{\bx}$ with velocity $\bv = \widehat{\bx} + v_y^0 \widehat{\mathbf{y}} + \widehat{\mathbf{z}}$.  The $y$-velocity is left variable so as to again study the trapped-passing boundary - at the initial particle location, the magnetic field is predominantly toroidal, making $v_y$ a reasonable proxy for $v_\parallel$.  A sample trajectory projected onto the $r$-$z$ plane with $v_y^0 = 0.7$ can be found in Figure \ref{fig:banana}.
Note that BFV incorrectly predicts a passing particle, while MI predicts an unphysical trajectory even after reducing the time-step by a factor of four relative to CBFV and BFV.  

\begin{figure}
	\centering
	\includegraphics[width=0.49\textwidth]{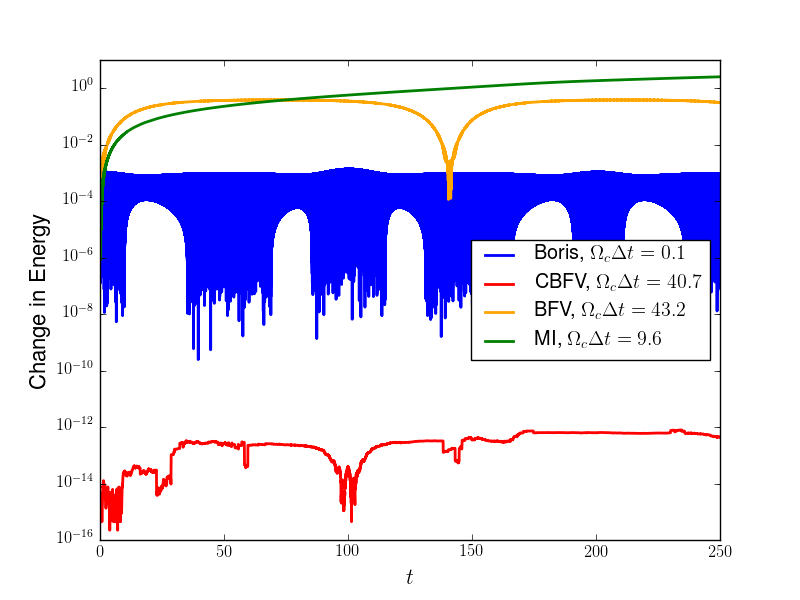}
	\includegraphics[width=0.49\textwidth]{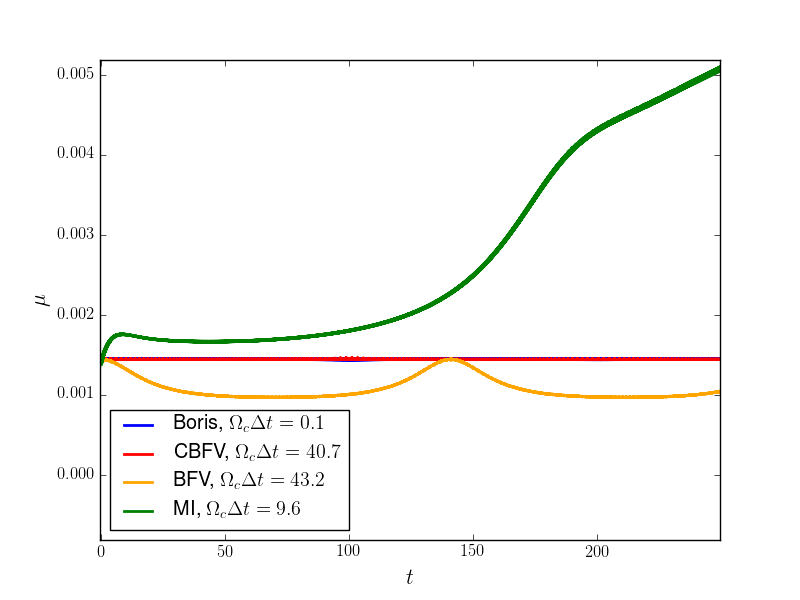}
	\caption{Energy error (left) and time history of magnetic moment (right) for the third test case in a Solov'ev equilbrium magnetic field with flux function electrostatic potential.}
	\label{fig:bananaEcons}
\end{figure}

We use a binary search to approximate the critical value of $v_y^0$ that separates the trapped and passing regimes.  We report the resulting values and percentage errors - the well-resolved Boris run is taken as the baseline in the absence of analytic theory - in Table \ref{table:bananabdry}.
\begin{table}[h]
\centering
\vspace{1em}
\caption{Critical values of $v_y^0$ that separate trapped and passing regimes found by each scheme.  Boris is taken as the reference case.  Values for MI are not reported due to the unphysical character of the orbits.}
\label{table:bananabdry}
\begin{tabular}{c | c c c}
	 & Boris ($\Oc \dt \ll 1$) & CBFV & BFV \\
	 \hline
	 Critical $v_y^0$ & 0.73107 & 0.73110 & 0.62725 \\
	 \% Error & $-$ & 0.004 & 14.2
\end{tabular}
\vspace{1em}
\end{table}
As in previous cases, the improved accuracy of the new method is attributable to improved energy and magnetic moment conservation, each of which is plotted in Figure \ref{fig:bananaEcons}.

\section{Conclusions}
In this work, we have introduced a first-of-kind numerical method for particle orbit integration that both (a) reproduces guiding center motion, including all first-order drifts, even when stepping over the gyration time-scale, and (b) conserves energy.  The time integrator is implicit, built upon the classical Crank-Nicolson algorithm, and is thus appropriate for incorporation into implicit PIC methods.  We have presented numerical results for single particle motion in several field configurations that demonstrate the dramatically improved conservation properties and accuracy of the scheme relative to earlier efforts.  

Opportunities for further development are myriad.  Practical implementation of the method in PIC codes stands to benefit tremendously from effective preconditioning of the nonlinear solve of the orbit integral.  Work in this direction, building on results for BFV in \cite{vu1995accurate} that analytically eliminate portions of the nonlinearity, is underway.  Additionally, the present scheme can only be said to capture the drift-kinetic limit - truly capturing the gyrokinetic limit requires removing the assumption that the gyroradius is small compared to length scales of interest.  Progress in this direction is of obvious interest.  Further refinement and generalization of the adaptive time-stepping strategy is also of importance for the scheme's future practical utility.

\section*{Acknowledgements}
This work was performed under the auspices of the U.S.\ Department of Energy by LLNL and LANL under contracts DE-AC52-07NA27344, DE-AC52-06NA25396, and supported by the Exascale Computing Project (17-SC-20-SC), a collaborative effort of the U.S. Department of Energy Office of Science and the National Nuclear Security Administration.  The authors wish to acknowledge valuable conversations regarding this work with Guangye Chen, William Taitano, C-S Chang, Joshua Burby, and Matthew Miecnikowski. 


\appendix
\section{Gyroaverage computations}
When computing the gyroaverages on the right of \eqref{eq:Gconstraint}, we may without loss of generality align $\bv_E$ with the $x$-axis.  Denoting gyrophase by $\gamma$ and writing $\bu = u ( \cos \gamma, \sin \gamma)^T$, we have
\begin{equation}
\begin{split}
	\left\langle \frac{1}{\left\| \bu + \bv_E \right\|^2} \right\rangle &= \frac{1}{2\pi u^2} \int_0^{2\pi} \frac{d\gamma}{1 + (v_E/u)^2 + 2 (v_E / u) \cos \gamma }, \\
	\left\langle \frac{\bu}{\left\| \bu + \bv_E \right\|^2} \right\rangle &= \frac{1}{2\pi u} \int_0^{2\pi} \frac{\left( \begin{array}{c} \cos \gamma \\ \sin \gamma \end{array} \right)d\gamma}{1 + (v_E/u)^2 + 2 (v_E / u) \cos \gamma } \\
	\left\langle \frac{\bu \bu}{\left\| \bu + \bv_E \right\|^2} \right\rangle &= \frac{1}{2\pi } \int_0^{2\pi} \frac{\left( \begin{array}{cc} \cos^2 \gamma & \sin \gamma \cos \gamma \\ \sin \gamma \cos \gamma & \sin^2 \gamma \end{array} \right)d\gamma}{1 + (v_E/u)^2 + 2 (v_E / u) \cos \gamma }.
\end{split}
\end{equation}
A computer algebra system can evaluate all of these integrals analytically.  We find
\begin{equation}
\begin{split}
	\left\langle \frac{1}{\left\| \bu + \bv_E \right\|^2} \right\rangle &= \frac{1}{u^2} \frac{1}{\left\lvert 1 - (v_E / u)^2 \right\rvert}, \\
	\left\langle \frac{\bu}{\left\| \bu + \bv_E \right\|^2} \right\rangle &= -\frac{v_E}{u^2} \frac{\eta^2}{\left\lvert 1 - (v_E / u)^2 \right\rvert} \widehat{\bx} \\
	\left\langle \frac{\bu \bu}{\left\| \bu + \bv_E \right\|^2} \right\rangle &= \eta^2 \left( \begin{array}{cc} \frac{1}{2} \frac{1 + (v_E/u)^2}{1 - (v_E/u)^2} & 0 \\ 0 & 1/2 \end{array} \right),
\end{split}
\end{equation}
where $\eta = \min \{ 1, u/v_E \}$ as in the main text.  Recalling that we assumed $\bv_E$ was in the $x$-direction, we can make these expressions general again with
\begin{equation}
\begin{split}
	\left\langle \frac{1}{\left\| \bu + \bv_E \right\|^2} \right\rangle &= \frac{1}{u^2} \frac{1}{\left\lvert 1 - (v_E / u)^2 \right\rvert}, \\
	\left\langle \frac{\bu}{\left\| \bu + \bv_E \right\|^2} \right\rangle &= -\frac{v_E}{u^2} \frac{\eta^2}{\left\lvert 1 - (v_E / u)^2 \right\rvert} \widehat{\bv}_E \\
	\left\langle \frac{\bu \bu}{\left\| \bu + \bv_E \right\|^2} \right\rangle &= \eta^2 \left( \frac{1}{2} \bI + \frac{(v_E/u)^2}{1 - (v_E/u)^2} \widehat{\bv}_E \widehat{\bv}_E \right).
\end{split}
\end{equation}
Here, again as in the main text, $\widehat{\bv}_E = \bv_E / v_E$.  

\section{Derivation of Time-step restrictions}
As noted in the main text, the fact that our scheme features a $\nabla B$ drift that differs from the true $\nabla B$ drift velocity induces an anomalous drift that (a) should be kept small and (b) should average to zeros on time-scales of interest.  Here, we derive the resulting restrictions on time-step size in the limits (i) $u^{n+1/2} \gg v_E$ and (ii) $u^{n+1/2} \ll v_E$.  

\subsection{The case $u^{n+1/2} \gg v_E$}
As noted in the main text - see \eqref{eq:Gperpudom} - in the case $u\nph > v_E$, the expression for $\bG$ simplified considerably:
\begin{equation}
	\bG = -\tilde{\mu} \left[ \bb (\nabla B)_\parallel + 2 (\nabla B)_\perp \right].
\end{equation}
Substituting this into \eqref{eq:goodProjec} , we find an expression for $\bv_{\nabla B}^\textrm{AP}$:
\begin{equation}
	\bv_{\nabla B}^\textrm{AP} = \frac{\bb}{\Omega_c} \times \frac{\tilde{\mu}}{m} \left\{ \frac{\mu_{\textrm{eff}}^\textrm{CN}}{\tilde{\mu}} \nabla B + \left( \bb - \frac{v_\parallel}{v_\perp} \widehat{\bv}_\perp \right) (\nabla B)_\parallel + 2 \left( \bI - \widehat{\bv}_\perp \widehat{\bv}_\perp \right) \cdot (\nabla B)_\perp \right\}.
\end{equation}

Recalling the relationship between $\mu$ and $\tilde{\mu}$ implied by \eqref{eq:mutildederivation} and subtracting the above from $\bv_{\nabla B}$, we find the following expression for the anomalous velocity:
\begin{equation}
	\Delta \bv_{\nabla B} = \frac{\bb}{\Oc} \times \frac{\tilde{\mu}}{m} \left\{ \frac{v_\parallel}{v_\perp} \widehat{\bv}_\perp (\nabla B)_\parallel + ( \bI - 2 \widehat{\bv}_\perp \widehat{\bv}_\perp ) \cdot (\nabla B)_\perp \right\}.
\end{equation}
At this point, it is useful to break $\Delta \bv_{\nabla B}$ into two pieces.  We write $\Delta \bv_{\nabla B} = \Delta \bv_{\nabla B}^1 + \Delta \bv_{\nabla B}^2$, with
\begin{equation}
	\Delta \bv_{\nabla B}^1 = \frac{\bb}{\Oc} \times \frac{\tilde{\mu}}{m} \frac{v_\parallel}{v_\perp} (\nabla B)_\parallel \widehat{\bv}_\perp, \qquad \Delta \bv_{\nabla B}^2 = \frac{\bb}{\Oc} \times \frac{\tilde{\mu}}{m} \left( \bI - 2 \widehat{\bv}_\perp \widehat{\bv}_\perp \right) \cdot (\nabla B)_\perp.
\end{equation}
Note that in the present limit in which gyration dominates the perpendicular velocity, each piece of $\Delta \bv_{\nabla B}$ has constant magnitude over time-scales short compared to variations in $\bB$ and $v_\parallel$.  The first piece simply rotates by angle $\theta$ given in \eqref{eq:thetaDef}.  In the second piece, the operator $(\bI - 2 \widehat{\bv}_\perp \widehat{\bv}_\perp) \cdot$ corresponds to a reflection about the $\widehat{\bv}_\perp$ axis.  Elementary geometry can thus be used to show that the second piece rotates by angle $2\theta$ at each time-step.  

As already noted, motion whose velocity has constant magnitude with velocity rotating by fixed angle transcribes a circle of radius given in \eqref{eq:discreteradius}.  Thus, the displacement induced by $\Delta v_{\nabla B}$ in this limit is a superposition of two circles of radius $R_1$ and $R_2$, respectively.  Elementary manipulation leads to formulae for these radii:
\begin{equation}
	R_1 = \frac{1}{8} \frac{\Oc^2 \dt^2}{\sqrt{ 1 + \Oc^2 \dt^2 / 4 }} \rho \delta_\parallel, \qquad R_2 = \frac{1}{16} \rho \delta _\perp \Oc^2 \dt^2, 
\end{equation}
where $\rho = u / \Oc$ is the gyroradius and $\delta_\parallel$ and $\delta_\perp$ are as defined in \eqref{eq:deltadefs}.  

To guarantee that the total anomalous displacement never exceeds the gyroradius, it thus suffices to impose $\max \{ R_1, R_2 \} \leq \rho / 2$.  This leads to the simultaneous time-step constraints
\begin{equation}
	\Oc \dt \leq \frac{\sqrt{2}}{\delta_\parallel} \left\{ 1 + \sqrt{1 + 4 \delta_\parallel^2} \right\}^{1/2}, \qquad \Oc \dt \leq 2 \sqrt{\frac{2}{\delta_\perp}}.
\end{equation}
Taking the minimum of these two restrictions in the limit $\delta_\parallel \ll 1$ gives the time-step restriction (ia) in Table \ref{table:dt_table}.  

We get the other restriction by analyzing the time required to traverse each circle and asking that it be less than the smallest time-scale of interest, denoted $\tau_\textrm{res}$.  The traversal time is simply the circumference of the circle - approximately $2 \pi R$ - over the anomalous drift speed.  For each of the two circles in question, we thus find traversal times
\begin{equation}
	\tau_\textrm{trav}^1 \approx \frac{4 \pi}{\Oc} \sqrt{ 1 + \Oc^2 \dt^2 / 4 }, \qquad \tau_\textrm{trav}^2 \approx \frac{\pi}{\Oc} \left( 1 + \Oc^2 \dt^2 / 4 \right).
\end{equation}
Insisting that both these quantities be smaller than $\tau_\textrm{res}$ leads to
\begin{equation}
	\Oc \dt \leq 2 \sqrt{ \frac{ \Oc \tau_\textrm{res}}{\pi} \min \left\{ 1, \frac{ \Oc \tau_\textrm{res} }{4 \pi} \right\} - 1 }.
\end{equation}
In the limit $\Oc \tau_\textrm{res} \gg 1$, this reduces to the restriction (ib) in Table \ref{table:dt_table}.  We note that no time-step satisfies this constraint if $\Oc \tau_\textrm{res} < \pi$.  However, in this case there is no separation in time-scales between gyration and $\tau_\textrm{res}$ and we are forced to resolve the gyration scale at any rate.  

\subsection{The case $u^{n+1/2} \ll v_E$}
We note that this limit corresponds to $\eta \ll 1$.  We thus keep only the terms proportional to $\eta^{-2}$ in evaluating the anomalous drift velocity in this case.  Using the general form of $\bG$ in \eqref{eq:genGdef}, we find that to leading order
\begin{equation}
	\Delta \bv_{\nabla B} = \frac{\bb}{\Oc} \times \frac{\tilde{\mu}}{m} \left\{ \left( \bI - \widehat{\bv}_\perp \widehat{\bv}_\perp \right) \cdot \widehat{\bv}_E \frac{2}{\eta^2} \left[ \widehat{\bv}_E \cdot (\nabla B)_\perp + \frac{v_\parallel}{v_\perp} (\nabla B)_\parallel \right] \right\}. 
\end{equation}
Next, we note that
\begin{equation} \label{eq:dvexp}
	\widehat{\bv}_\perp \cdot \widehat{\bv}_E \approx \frac{\left( \bv_E + \bu \right) \cdot \widehat{\bv}_E }{\left\| \bv_E + u \right\| } \approx 1 + \eta \cos \psi,
\end{equation}
where in the second approximation we've again assumed $u \ll v_E$, and $\cos \psi = \widehat{\bu} \cdot \widehat{\bv}_E$ defines $\psi$ as the angle between $\bu$ and $\bv_E$.  This leads us to the conclusion
\begin{equation}
	\left( \bI - \widehat{\bv}_\perp \widehat{\bv}_\perp \right) \cdot \widehat{\bv}_E \approx \widehat{\bv}_E - \widehat{\bv}_\perp (1 + \eta \cos \psi) \approx - \eta \left( \widehat{\bu} + \widehat{\bv}_E \cos \psi \right)
\end{equation}
to leading order in $\eta$, where we've used $\widehat{\bv}_\perp - \widehat{\bv}_E \approx \eta \widehat{\bu}$ in the second approximation.  

Substituting this into \eqref{eq:dvexp}, we have
\begin{equation}
	\Delta \bv_{\nabla B} \approx -\frac{2}{\eta} \frac{\tilde{\mu}}{m} \frac{\bb}{\Oc} \times \left( \widehat{\bu} + \widehat{\bv}_E \cos \psi \right) \left[ \widehat{\bv}_E \cdot (\nabla B)_\perp + \frac{v_\parallel}{v_\perp} (\nabla B)_\parallel \right].
\end{equation}
Note that $\widehat{\bu}$ rotates by $\theta$ at every time-step, and the angle $\psi$ changes by $\theta$ as well so long as $\bv_E$ changes little in a time-step.  Thus, for $\Oc \dt \gg 1$ - i.e.\ $\theta \approx \pi$ - $\Delta \bv_{\nabla B}$ approximately flips sign at every time-step.  More generally, for $\theta$ near but not equal to $\pi$, we my reasonably estimate that the anomalous displacement averages to zero over $2 \pi / \theta$ time-steps.  As such, we may bound the maximum anomalous displacement by its value after $\pi / \theta$ time-steps, and insist that this value be less than $\rho$.  That is, we insist that $\| \Delta \bv_{\nabla B} \| \dt \pi / \theta \leq \rho$.  For large $\Oc \dt$, Taylor expansion shows that $\theta \approx \pi - 4/\Oc \dt$.  Taking the worst case scenario in which $\| \widehat{\bu} - \widehat{\bv}_E \cos \psi \| = 2$ and $\nabla B$ is parallel to $\bv_E$, and recalling the subtlety that the relevant $u$ in the definition of $\eta$ is $u\nph$ while in $\mu$ it is $u^n$, this constraint reduces to
\begin{equation}
	2 \left( \delta_E + \delta_\parallel \right) \frac{\Oc^3 \dt^3 / 4}{\sqrt{1 + \Oc^2 \dt^2 / 4}} \left(\frac{\pi}{\pi - 4/\Oc \dt} \right) \leq 1
\end{equation}
Again assuming $\Oc \dt \gg 1$ gives the constraint (iia) in Table \ref{table:dt_table}.

The constraint on the time-scale over which the anomalous displacement averages to zero is in this case simply $\pi \dt / \theta \leq \tau_\textrm{res}$.  Again in the limit $\Oc \dt \gg 1$, this reduces to constraint (iib) in Table \ref{table:dt_table}.
\bibliographystyle{plain}
\bibliography{APbib}

\begin{thebibliography}{10}

\bibitem{birdsall2004plasma}
C.K. Birdsall and A.B. Langdon.
\newblock {\em Plasma physics via computer simulation}.
\newblock CRC press, 2004.

\bibitem{brackbill1985simulation}
J.U. Brackbill and D.W. Forslund.
\newblock Simulation of low-frequency, electromagnetic phenomena in plasmas.
\newblock In {\em Multiple time scales}, pages 271--310. Elsevier, 1985.

\bibitem{cary:2009aa}
John~R. Cary.
\newblock Hamiltonian theory of guiding-center motion.
\newblock {\em Reviews of Modern Physics}, 81(2):693--738, 2009.

\bibitem{cerfon2010one}
A.J. Cerfon and J.P. Freidberg.
\newblock ``one size fits all'' analytic solutions to the {G}rad--{S}hafranov
  equation.
\newblock {\em Physics of Plasmas}, 17(3):032502, 2010.

\bibitem{chacon2013charge}
L.~Chac{\'o}n, G.~Chen, and D.C. Barnes.
\newblock A charge-and energy-conserving implicit, electrostatic
  particle-in-cell algorithm on mapped computational meshes.
\newblock {\em Journal of Computational Physics}, 233:1--9, 2013.

\bibitem{chang2004numerical}
C.-S. Chang, S.~Ku, and H.~Weitzner.
\newblock Numerical study of neoclassical plasma pedestal in a tokamak
  geometry.
\newblock {\em Physics of Plasmas}, 11(5):2649--2667, 2004.

\bibitem{chen2014energy}
G.~Chen and L.~Chac{\'o}n.
\newblock An energy-and charge-conserving, nonlinearly implicit,
  electromagnetic 1{D}-3{V} {V}lasov--{D}arwin particle-in-cell algorithm.
\newblock {\em Computer Physics Communications}, 185(10):2391--2402, 2014.

\bibitem{chen2015multi}
G.~Chen and L.~Chac{\'o}n.
\newblock A multi-dimensional, energy-and charge-conserving, nonlinearly
  implicit, electromagnetic {V}lasov--{D}arwin particle-in-cell algorithm.
\newblock {\em Computer Physics Communications}, 197:73--87, 2015.

\bibitem{chen2011energy}
G.~Chen, L.~Chac{\'o}n, and D.C. Barnes.
\newblock An energy-and charge-conserving, implicit, electrostatic
  particle-in-cell algorithm.
\newblock {\em Journal of Computational Physics}, 230(18):7018--7036, 2011.

\bibitem{chen2012efficient}
G.~Chen, L.~Chac{\'o}n, and D.C. Barnes.
\newblock An efficient mixed-precision, hybrid {CPU}--{GPU} implementation of a
  nonlinearly implicit one-dimensional particle-in-cell algorithm.
\newblock {\em Journal of Computational Physics}, 231(16):5374--5388, 2012.

\bibitem{chen2014fluid}
G.~Chen, L.~Chac{\'o}n, C.A. Leibs, D.A. Knoll, and W.~Taitano.
\newblock Fluid preconditioning for {N}ewton--{K}rylov-based, fully implicit,
  electrostatic particle-in-cell simulations.
\newblock {\em Journal of computational physics}, 258:555--567, 2014.

\bibitem{chen2001gyrokinetic}
Y.~Chen and S.~Parker.
\newblock Gyrokinetic turbulence simulations with kinetic electrons.
\newblock {\em Physics of Plasmas}, 8(5):2095--2100, 2001.

\bibitem{cohen2007large}
R.H. Cohen, A.~Friedman, D.P. Grote, and J.-L. Vay.
\newblock Large-timestep mover for particle simulations of arbitrarily
  magnetized species.
\newblock {\em Nuclear Instruments and Methods in Physics Research Section A:
  Accelerators, Spectrometers, Detectors and Associated Equipment},
  577(1-2):52--57, 2007.

\bibitem{ethier2005gyrokinetic}
S.~Ethier, W.M. Tang, and Z.~Lin.
\newblock Gyrokinetic particle-in-cell simulations of plasma microturbulence on
  advanced computing platforms.
\newblock In {\em Journal of Physics: Conference Series}, volume~16, page~1.
  IOP Publishing, 2005.

\bibitem{filbet2016asymptotically}
F.~Filbet and L.M. Rodrigues.
\newblock Asymptotically stable particle-in-cell methods for the
  vlasov--poisson system with a strong external magnetic field.
\newblock {\em SIAM Journal on Numerical Analysis}, 54(2):1120--1146, 2016.

\bibitem{filbet2017asymptotically}
F.~Filbet and L.M. Rodrigues.
\newblock Asymptotically preserving particle-in-cell methods for inhomogeneous
  strongly magnetized plasmas.
\newblock {\em SIAM Journal on Numerical Analysis}, 55(5):2416--2443, 2017.

\bibitem{FIVAZ199827}
M.~Fivaz, S.~Brunner, G.~de~Ridder, O.~Sauter, T.M. Tran, J.~Vaclavik,
  L.~Villard, and K.~Appert.
\newblock Finite element approach to global gyrokinetic particle-in-cell
  simulations using magnetic coordinates.
\newblock {\em Computer Physics Communications}, 111(1):27 -- 47, 1998.

\bibitem{genoni2010fast}
T.C. Genoni, R.E. Clark, and D.R. Welch.
\newblock A fast implicit algorithm for highly magnetized charged particle
  motion.
\newblock {\em Open Plasma Physics Journal}, 3:36--41, 2010.

\bibitem{hazeltine2003plasma}
R.D. Hazeltine and J.D. Meiss.
\newblock {\em Plasma confinement}.
\newblock Courier Corporation, 2003.

\bibitem{hazeltine2018framework}
R.D. Hazeltine and F.~Waelbrock.
\newblock {\em The framework of plasma physics}.
\newblock CRC Press, 2018.

\bibitem{Krommes_2007}
J.A. Krommes.
\newblock Nonequilibrium gyrokinetic fluctuation theory and sampling noise in
  gyrokinetic particle-in-cell simulations.
\newblock {\em Physics of Plasmas}, 14(9):090501, Sep 2007.

\bibitem{parker1991numerical}
S..E. Parker and C.K. Birdsall.
\newblock Numerical error in electron orbits with large $\omega_{ce} \delta t$.
\newblock {\em Journal of Computational Physics}, 97(1):91--102, 1991.

\bibitem{pataki2013fast}
A.~Pataki, A..J Cerfon, J.P. Freidberg, L.~Greengard, and M.~O'Neil.
\newblock A fast, high-order solver for the {G}rad--{S}hafranov equation.
\newblock {\em Journal of Computational Physics}, 243:28--45, 2013.

\bibitem{simo1992exact}
J.C. Simo, N.~Tarnow, and K.K. Wong.
\newblock Exact energy-momentum conserving algorithms and symplectic schemes
  for nonlinear dynamics.
\newblock {\em Computer methods in applied mechanics and engineering},
  100(1):63--116, 1992.

\bibitem{taitano2013development}
W.T. Taitano, D.A. Knoll, L.~Chac{\'o}n, and G.~Chen.
\newblock Development of a consistent and stable fully implicit moment method
  for {V}lasov--{A}mp{\`e}re particle in cell ({PIC}) system.
\newblock {\em SIAM Journal on Scientific Computing}, 35(5):S126--S149, 2013.

\bibitem{vu1995accurate}
H.X. Vu and J.U. Brackbill.
\newblock Accurate numerical solution of charged particle motion in a magnetic
  field.
\newblock {\em Journal of Computational Physics}, 116(2):384--387, 1995.

\end{thebibliography}

\end{document}